\documentclass[
onecolumn;
superscriptaddress,
showpacs,
preprintnumbers,
amsmath,
amssymb,
aps,
prd,
]{revtex4-2}

\usepackage{graphicx}
\usepackage{dcolumn}
\usepackage{bm}

\begin{document}

\title{Non-self-dual nontopological soliton in a pure Chern-Simons gauge model}

\author{A.~Yu.~Loginov}

\email{a.yu.loginov@tusur.ru}

\affiliation{Laboratory of Applied Mathematics and Theoretical Physics, Tomsk State
             University of Control Systems and Radioelectronics, 634050 Tomsk, Russia}


\begin{abstract}
A nontopological soliton of the Q-ball type in a Chern-Simons-Higgs gauge model
is studied using both analytical and numerical methods.
The      general       non-self-dual         case          is       considered.
It is shown that the soliton solution  is  an extremum of the energy functional
at a fixed Noether charge.
A differential relation between  the  energy,  Noether charge, and the boundary
value of the gauge potential of the soliton is derived.
A linear relation between the components  of  the   soliton energy is obtained.
The parametric domain of existence  of  the  soliton  solution  is  determined.
It is established that the soliton properties  depend significantly on the form
of the self-interaction potential of the scalar field.
In particular, the energy and charge  of the soliton can take arbitrarily large
values only if  the  self-interaction potential has two degenerate zero minima.
\end{abstract}

\maketitle

\section{Introduction} \label{sec:I}

In  odd-dimensional  space-time,  gauge  models   may   include  a  topological
Chern-Simons term  in  addition  to  the  kinetic Maxwell (or Yang-Mills) term.
Such gauge  models  were considered  in \cite{schonfeld_npb_1981, djt_prl_1982,
 djt_ann_phys_1982} for the $(2 + 1)$-dimensional case, in which the derivative
part of  the   Chern-Simons   term     is    quadratic   in  the  gauge fields.
It was shown in these works that the Chern-Simons term  gives rise to a massive
gauge field without breaking the gauge symmetry.
Note also that the  Chern-Simons term is induced dynamically in the presence of
fermions and  can  be  considered  as  a  polarization  effect of the fermionic
vacuum \cite{redlich_prl_1984, redlich_prd_1984}.

When  only   the   Maxwell   term   is   present   in   the   Lagrangian   of a
$(2+1)$-dimensional Abelian gauge model, its soliton solutions are electrically
neutral and  have zero  angular momentum, since otherwise their energy would be
infinite.
Adding the  Chern-Simons   term   to   the  Lagrangian  drastically  alters the
field   dynamics,    making     possible    the   existence   of  finite-energy
soliton solutions (topological, nontopological, or both)  with nonzero electric
charge    and    angular   momentum   \cite{paul_plb_1986, inozemtsev_epl_1988,
 lozano_prd_1988, khare_plb_1989,  lee_plb_1990, khare_plb_1991, kimm_prd_1996,
 deshaies_prd_2006,   bazeia_prd_2012,   loginov_jetp_2014,   navarro_prd_2017,
 samoilenka_prd_2017,  navarro_prd_2019, andrade_prd_2020, ivashkin_2025}.

The Maxwell term is  of  second  order  in  space-time derivatives, whereas the
Chern-Simons term is of first order.
Hence, the Chern-Simons  term     dominates   over   the   Maxwell  term in the
long-wavelength limit, playing an important  role in various areas of condensed
matter physics.
In this limit, the Maxwell term becomes  irrelevant  and  can  be  dropped from
the  Lagrangian,   resulting    in    a    pure    Chern-Simons    gauge theory
\cite{hagen_ann_phys_1984, hagen_prd_1985}.
Similar to  the  Maxwell-Chern-Simons  gauge  models,  pure  Chern-Simons gauge
models may have topological, nontopological, or both types of soliton solutions
\cite{hong_prl_1990,      jw_prl_1990,      jlw_prd_1990,      jatkar_plb_1990,
 bazeia_prd_1991,    bazeia_prd_1991_b,    torres_prd_1992,     ghosh_plb_1996,
 arthur_prd_1996,  bazeia_plb_2017}.
As in the previous case,  these  solitons  are  electrically  charged  and have
nonzero angular momentum.
A comprehensive  review  of   Chern-Simons   gauge   models  and  their soliton
solutions can be found in \cite{jackiw_ptps_1992, dunne_1995, horvathy_pr_2009}.

Topological solitons (vortices) of gauge models with the Chern-Simons term have
not only quantized magnetic flux but also quantized electric charge and angular
momentum.
In the framework of quantum field  theory,  these  vortices  are interpreted as
anyons, particles with fractional electric charge and angular momentum.
It is expected that these particles  may  play a crucial role in the fractional
quantum   Hall   effect   \cite{prange_1990}    and   the   high-$T_{\text{c}}$
superconductivity \cite{wilczek_1990}.

One of  the  field  components of  the  aforementioned  soliton  solutions is a
self-interacting scalar field.
For a certain form of the self-interaction  potential,  the energy of a soliton
with a given  magnetic  flux  reaches  the  minimum  possible value, called the
Bogomol'nyi bound \cite{bogomolny_sjnp_1976}.
In this case, the  soliton fields satisfy both the second-order field equations
and the first-order self-dual equations.
Soliton solutions of the first-order  self-dual equations  are more amenable to
analytical study than those of the second-order field equations.
As a result,  self-dual  soliton  solutions  are  the  main object of research.
However, solitons of  Chern-Simons  gauge  models  can  exist  not  only at the
self-dual parametric  point  but  also  in  a  parametric  region  in which the
self-duality condition is not satisfied.

In this work, we study a nontopological soliton of the Chern-Simons-Higgs gauge
model in the non-self-dual parametric region.
We establish  the  boundaries  of  this  parametric  region and  derive several
important properties of the nontopological soliton.
We find that in the general case, corresponding   to the absence of spontaneous
symmetry breaking, the  energy   and   electric charge of the soliton cannot be
arbitrarily large.
In contrast, they can take arbitrarily large values on the part of the boundary
of the parametric  region  where  the  gauge  symmetry is spontaneously broken.
Note that the self-dual nontopological soliton of the Chern-Simons-Higgs  model
considered in \cite{jlw_prd_1990} corresponds to a critical boundary point, and
its energy and charge can take arbitrarily large values.
A number of issues concerning the non-self-dual nontopological  solitons of the
Chern-Simons-Higgs   model   were   considered    in    \cite{ bazeia_prd_1991,
 bazeia_prd_1991_b}.

This  paper  is  structured as follows.
In Sec.~\ref{sec:II}, we  briefly  describe  the  Lagrangian, symmetries, field
equations, and some properties of the model under consideration.
In Sec.~\ref{sec:III}, the general  properties  of  the  nontopological soliton
are derived and discussed.
In Sec.~\ref{sec:V}, we present and discuss the results of a numerical study of
the soliton.
In the final section, we  briefly summarize the results obtained in this paper.
In Appendix A, we discuss the  behaviour  of  the  nontopological  soliton in a
limiting regime.

Throughout the paper, the natural units $c = 1$ and $\hbar = 1$ are used.

\section{The model}   \label{sec:II}

The  Chern-Simons-Higgs  model  is  described  by  the  action
\begin{equation}
S=\int d^{3}x\left[ \frac{\kappa }{2}\epsilon ^{\alpha \beta \gamma
}A_{\alpha }\partial_{\beta}A_{\gamma}+\left\vert D_{\mu }\phi \right\vert
^{2}-V\left( \left\vert \phi \right\vert \right) \right],          \label{II:1}
\end{equation}
where the covariant derivative $D_{\mu}\phi =\partial_{\mu}\phi+ieA_{\mu}\phi$,
the metric tensor  $g_{\mu \nu}  =  \text{diag}(1,-1,-1)$,  and the Levi-Civita
symbol $\epsilon^{\alpha \beta \gamma}$ is  defined according to the convention
$\epsilon^{0 1 2} = \epsilon_{0 1 2} = 1$.
In eq.~\eqref{II:1},  $\kappa$  is  a  coupling  constant  that  determines the
strength of the Chern-Simons term $\mathcal{L}_{CS}=2^{-1}\kappa\epsilon^{\alpha
\beta \gamma }A_{\alpha }\partial_{\beta}A_{\gamma}$, while the electromagnetic
coupling constant  $e$  determines  the  strength  of  the  minimal interaction
between the gauge field $A_{\mu}$ and the complex scalar field $\phi$.
The self-interaction potential of the complex scalar field $\phi$ has  the form
\begin{equation}
V\left( \left\vert \phi \right\vert \right) =\frac{m^{2}\left\vert \phi
\right\vert^{2}}{1+\varepsilon ^{2}}\left[\left(1 - v^{-2}\left\vert \phi
\right\vert^{2}\right)^{2} + \varepsilon^{2}\right].               \label{II:2}
\end{equation}
The sixth-order potential \eqref{II:2} is  renormalizable in $(2+1)$ dimensions
and  is   invariant   under   the   $U(1)$   transformations  $\phi \rightarrow
\exp(-i\Lambda) \phi$.
This  form   of   writing    of    a   sixth-order   potential   was   used  in
\cite{lee_pang_1992}; its advantage  is  that the potential \eqref{II:2}  has a
zero global minimum at $\phi = 0$  for  any  values of the real parameters $m$,
$v$, and $\varepsilon$.

By varying the action \eqref{II:1} with respect  to  the  corresponding fields,
we obtain the field equations of the model,
\begin{eqnarray}
D_{\mu }D^{\mu }\phi  &=&-\frac{\partial V}{\partial\phi^{\ast}}, \label{II:3a}
 \\
\frac{\kappa}{2}\epsilon^{\mu\nu\sigma}F_{\nu\sigma }&=&j^{\mu},  \label{II:3b}
\end{eqnarray}
where the conserved matter  current  density  $j^{\mu} = (\rho, \mathbf{j})$ is
\begin{equation}
j_{\mu}=ie\left[\phi^{\ast}D_{\mu}\phi-\phi \left(D_{\mu}\phi
\right) ^{\ast }\right].                                           \label{II:4}
\end{equation}
The field equations \eqref{II:3a}--\eqref{II:4} are invariant  under the  gauge
transformations
\begin{equation}
\phi \left( x\right) \rightarrow \phi ^{\prime }\left( x\right) =\exp \left(
-i\Lambda \left( x\right) \right) \phi \left( x\right) ,\;A_{\mu }\left(
x\right) \rightarrow A_{\mu }^{\prime }\left( x\right) = A_{\mu }\left(
x\right) + e^{-1} \partial_{\mu}\Lambda \left( x\right).           \label{II:5}
\end{equation}
In contrast,  the  Lagrangian  density  entering  into  eq.~\eqref{II:1} is not
invariant under the gauge transformation  \eqref{II:5} as the Chern-Simons term
changes by a total derivative: $\mathcal{L}_{CS} \rightarrow \mathcal{L}_{CS} +
\kappa \left( 2e\right)^{-1}\partial_{\alpha } \left( \Lambda \epsilon ^{\alpha
\beta \gamma }\partial_{\beta}A_{\gamma }\right)$.
Nevertheless, the  action  \eqref{II:1}  remains gauge invariant, provided that
the gauge parameter $\Lambda(x)$ tends to  zero sufficiently fast at space-time
infinity.
As for the discrete transformations,  the  Lagrangian  density  \eqref{II:1} is
not invariant under the $P$ and $T$-transformations, since the Chern-Simon term
$\mathcal{L}_{CS}$ changes sign in this case.
At the same  time,  it  is  invariant  under  the $C$ and $PT$-transformations.

Since the model under consideration is a pure Chern-Simons model, and therefore
does not contain the Maxwell term,  the  field equation \eqref{II:3b} is of the
first order.
It implies  Gauss's law
\begin{equation}
B = -\kappa^{-1}\rho                                               \label{II:6}
\end{equation}
and Amp{\`e}re's law
\begin{equation}
E^{i} = \kappa^{-1}\epsilon^{ij} j^{j},                            \label{II:7}
\end{equation}
of the Chern-Simons electrodynamics, where  the  magnetic field  strength  $B =
-F_{12} = F_{21}$,  the electric  field strength $E^{i} = F^{i0} = F_{0i}$, and
we use the spatial antisymmetric symbol $\epsilon^{ij}$ with $\epsilon^{12}=1$.
Written in the form $j^{i} = -\kappa \epsilon^{i j} E^{j}$, eq.~\eqref{II:7} is
precisely Hall's law with the Hall conductance $\sigma_{H} = -\kappa$.

Eq.~\eqref{II:6} tells  us  that in the model~\eqref{II:1}, the electric charge
and magnetic flux of a localized field  configuration  are linearly  related as
\begin{equation}
Q = -\kappa \Phi.                                                  \label{II:8}
\end{equation}
We can also use eq.~\eqref{II:6} to express the time component $A_{0}$ in terms
of the magnetic  field   strength  $B$  and  the  complex  scalar  field  $\phi
= \left\vert\phi \right\vert e^{i \arg \left( \phi \right)}$ as
\begin{equation}
A_{0}=\frac{\kappa}{2e^{2}}\frac{B}{\left\vert\phi\right\vert^{2}}-\frac{
1}{e}\partial _{0}\arg \left( \phi \right).                        \label{II:9}
\end{equation}
In turn, eq.~\eqref{II:7} implies  that the matter current density $\mathbf{j}$
is perpendicular to the electric field strength $\mathbf{E}$.

By varying the curved-space form of the action  \eqref{II:1}  in the metric and
using the   expression   $\delta S   =  2^{-1} \int d^{3}x \sqrt{-g}T_{\mu \nu}
\delta g^{\mu \nu }$, we obtain  the  symmetric  energy-momentum  tensor of the
model,
\begin{equation}
T_{\mu \nu }=\left( D_{\mu }\phi \right) ^{\ast }D_{\nu }\phi +D_{\mu }\phi
\left(D_{\nu}\phi\right)^{\ast}-g_{\mu \nu}\left[\left\vert D_{\sigma
}\phi\right\vert^{2}-V\left(\left\vert \phi \right\vert\right)\right].
                                                                  \label{II:10}
\end{equation}
It follows from eqs.~\eqref{II:1}  and \eqref{II:10} that the Chern-Simons term
$S_{CS} = 2^{-1}  \kappa \int d^{3}x \epsilon ^{\alpha \beta \gamma} A_{\alpha}
\partial_{\beta}A_{\gamma}$  does not contribute to the energy-momentum tensor.
This  is  because $S_{CS}$  has  a  generally  covariant  form,  and  therefore
preserves it in a curved space-time.
Since  $S_{CS}$  does  not  depend  on  the  metric  tensor  $g^{\mu \nu}$, the
variational   derivative    $\delta S_{CS}/\delta g^{\mu \nu }$   vanishes, and
therefore $S_{CS}$ does not contribute to $T_{\mu \nu}$ directly.
Instead,  $S_{CS}$  influences  $T_{\mu \nu}$   indirectly,  since the coupling
constant $\kappa$ is included in the field equation \eqref{II:3b}.

Later on, we shall need the expression  for the energy of a field configuration
of the model, 
\begin{eqnarray}
E &=&\int d^{2}x\left[ \left\vert D_{0}\phi \right\vert ^{2} + \left\vert
\mathbf{D}\phi \right\vert ^{2}+V\left( \left\vert \phi \right\vert \right)
\right]  \nonumber
  \\
&=&\int d^{2}x\left[ \left( \partial _{0}\left\vert \phi \right\vert \right)
^{2}+\frac{\kappa ^{2}}{4e^{2}}\frac{B^{2}}{\left\vert \phi \right\vert^{2}}
+\left\vert \mathbf{D}\phi \right\vert ^{2}+V\left( \left\vert \phi
\right\vert \right) \right],                                      \label{II:11}
\end{eqnarray}
where eq.~\eqref{II:9}  has   been   used   to   obtain   the   second  line of
eq.~\eqref{II:11}.
Eq.~\eqref{II:11} tells us that the magnetic field strength $B$ vanishes at the
zeroes of the scalar field  $\phi$,  which  is  a  consequence  of  Gauss's law
\eqref{II:6}.
We shall  also  need  the  expression  for  the  angular  momentum  of  a field
configuration of the model,
\begin{equation}
J=\int d^{2}x\left[ x^{1}T^{02}-x^{2}T^{01}\right] =-\int d^{2}x\,\mathbf{
x\times}\!\left[ \left( D_{0}\phi \right)^{\ast }\mathbf{D}\phi +D_{0}\phi
\,\left(\mathbf{D}\phi \right)^{\ast}\right].                     \label{II:12}
\end{equation}

As noted  above,  the  potential $V\left(\left\vert \phi \right\vert\right)$ in
eq.~\eqref{II:2} has a zero global  minimum  at  the symmetric point $\phi = 0$
for  any  values  of the parameters  $m$, $v$, and $\varepsilon$.
In addition, the potential $V\left(\left\vert \phi \right\vert\right)$ also has
degenerate  local  asymmetric  minima  on  the  circle  $\left\vert \phi \right
\vert_{\min}=3^{-1/2}v\bigl(2+\left(1-3\varepsilon^{2} \right)^{1/2}\bigr)^{1/2
}$, provided that the parameter $\varepsilon \in \left[0, 3^{-1/2}\right)$.
The  value  of  $V\left(\left\vert \phi \right\vert_{\min}\right)$  vanishes at
$\varepsilon = 0$,  with  the  result  that the asymmetric and symmetric minima
become degenerate.
In this case, the model has the symmetric unbroken vacuum at $\phi = 0$ and the
asymmetric broken vacuum at $\left\vert \phi \right\vert=\left\vert \phi \right
\vert_{\min}$.

The implementation of the Higgs  mechanism  in the pure Chern-Simons case has a
number of features \cite{deser_yang_mpla_1989, wen_zee_jphysfr_1989}.
In the unbroken vacuum, the gauge  field  is nonpropagating, and thus there are
only two propagating real scalar modes of mass $m$.
In the broken vacuum, a  massless  component of the scalar field (the Goldstone
boson) combines with  the  longitudinal  part  of  the gauge field to produce a
massive gauge mode.
As a result, in the broken vacuum there is one massive gauge mode of mass $m_{A
}$, and one real massive scalar mode (the Higgs boson) of mass $m_{H}$, where
\begin{equation}
m_{A} = 2e^{2}v^{2}\kappa^{-1} \quad \text{and} \quad m_{H} = 2 m.\label{II:14}
\end{equation}
Note that in contrast  to  the conventional Higgs  mechanism, in which the mass
of a gauge mode  is  proportional  to  the magnitude of the scalar field vacuum
expectation value, the  mass  $m_{A}$  is  proportional  to  the  square of the
magnitude of the scalar field vacuum expectation value.
The reason  is  that   the   Chern-Simons  term  is   first-order  in spacetime
derivatives, whereas the Maxwell term is second-order.

\section{The nontopological soliton and its properties} \label{sec:III}

In general, the masses $m_{A}$ and $m_{H}$  are  independent and not related to
each other.
In the self-dual case, the  masses  $m_{A}$  and  $m_{H}$ satisfy the condition
$m_{A} = m_{H} = 2 m$.
In this exceptional case,  static  field  configurations  of  the model satisfy
both field  equations  \eqref{II:3a}--\eqref{II:3b},  one of which is of second
order, and   the  self-dual  equations  of  first  order   \cite{hong_prl_1990,
 jw_prl_1990}.
In refs.~\cite{hong_prl_1990,  jw_prl_1990,  jlw_prd_1990},  it  was shown that
in the  self-dual case,  the  model \eqref{II:1} possesses both topological and
nontopological  soliton  solutions,  and their properties are studied.

Let us define one  more dimensionless parameter $\tau = m_{A}/\left(2m\right) =
e^{2}v^{2}/\left(m \kappa \right)$ in addition  to  $\varepsilon$.
The self-dual case corresponds  to the values $\varepsilon = 0$ and $\tau = 1$.
In this work,  we  study  nontopological  solitons  of  the  model \eqref{II:1}
corresponding to  the  non-self-dual  case of  $\tau \ne 1$  with $\varepsilon$
being either zero or nonzero.
It is easy  to  see  that  topological  solitons  (vortices)  do  not  exist if
$\varepsilon \ne 0$.
This is  because  the  potential  \eqref{II:2}  vanishes   only  at  the  point
$\phi = 0$ in this case.   
It follows that  for  a  finite  energy  field  configuration, the scalar field
$\phi$ must tend to zero at spatial infinity.
However, such  a  boundary  condition  is  topologically  trivial,  which makes
the existence of topological solitons impossible.

To describe a nontopological soliton, we shall use the axially symmetric ansatz
\begin{align}
\phi \left( r,\theta ,t\right) & =v\exp \left( -i\omega t\right) f\left(
r\right),                                                           \nonumber
 \\
A^{\mu }\left( r,\theta \right) & =\left[ a_{0}\left( r\right) ,
\epsilon ^{ij}n^{j}a\left( r\right)/(e r) \right]\!,              \label{III:2}
\end{align}
where   $\epsilon^{ij}$   is    the    two-dimensional    antisymmetric  tensor
($\epsilon^{12}=1$) and $\mathbf{n}=\left[\cos(\theta), \sin(\theta)\right]$ is
the radial unit vector.
The ansatz \eqref{III:2}  retains  its  form  under  the  gauge transformations
\eqref{II:5} with $\Lambda = \Delta\omega\,t$.
To fix  the  gauge,  we  impose  the  boundary  condition  $a_{0}(\infty) = 0$.
Substituting    ansatz     \eqref{III:2}     into     the     field   equations
\eqref{II:3a}--\eqref{II:3b},  we  obtain  the  system of ordinary differential
equations for the ansatz functions:
\begin{eqnarray}
f^{\prime \prime }+\frac{f^{\prime }}{r}+\Omega ^{2}f-\frac{a^{2}}{r^{2}}f-%
\frac{1}{2}\frac{\partial \tilde{V}}{\partial f} &=&0,           \label{III:3a}
 \\
a^{\prime }-m_{A}r\Omega f^{2} &=&0,                             \label{III:3b}
 \\
\Omega ^{\prime }-m_{A}\frac{af^{2}}{r} &=&0,                    \label{III:3c}
\end{eqnarray}
where the combination $\Omega = \omega - e a_{0}$,  and  the rescaled potential
\begin{equation}
\tilde{V}\left( f\right) =v^{-2}V\left( f\right) =\frac{m^{2}f^{2}}{
1+\varepsilon ^{2}}\left( 1+\varepsilon ^{2}-2f^{2}+f^{4}\right). \label{III:4}
\end{equation}
Note that, unlike  eq.~\eqref{III:3a}, eqs.~\eqref{III:3b}  and  \eqref{III:3c}
(Gauss's law and Amp{\`e}re's law, respectively) are of first order, which is a
characteristic  property  of  a  pure Chern-Simons gauge theory.
The two first order equations \eqref{III:3b} and \eqref{III:3c}  are equivalent
to the single second-order equation for $\Omega$,
\begin{equation}
\Omega ^{\prime \prime }+\left( \frac{1}{r}-2\frac{f^{\prime }}{f}\right)
\Omega ^{\prime }-m_{A}^{2}f^{4}\Omega = 0,                       \label{III:5}
\end{equation}
or for $a$,
\begin{equation}
a^{\prime \prime }-\left( \frac{1}{r}+2\frac{f^{\prime }}{f}\right)
a^{\prime }-m_{A}^{2}f^{4}a=0.                                    \label{III:6}
\end{equation}
We see that if  $(f, \Omega, a)$  is a solution to the system \eqref{III:3a} --
\eqref{III:3c},  then  $(f, -\Omega, -a)$  is  also  a  solution, which results
from the $C$-invariance of the model \eqref{II:1}.
We also see that the ansatz function $f$ is defined up to sign, since the model
\eqref{II:1}   is    invariant    under   the   global   gauge   transformation
$\phi \rightarrow -\phi$.

Substituting the  ansatz  \eqref{III:2}  into eqs.~\eqref{II:4}, \eqref{II:11},
and \eqref{II:12}, we obtain the expressions for the electric charge,
\begin{equation}
Q=4\pi ev^{2}\int\nolimits_{0}^{\infty }\Omega f^{2}rdr,          \label{III:7}
\end{equation}
the energy,
\begin{equation}
E=2 \pi v^{2}\int\nolimits_{0}^{\infty}\left[f^{\prime 2}+\Omega^{2}f^{2}+
\frac{a^{2}}{r^{2}}f^{2}+\tilde{V}\right] rdr,                    \label{III:8}
\end{equation}
and the angular momentum
\begin{equation}
J=4\pi v^{2}\int\nolimits_{0}^{\infty} a\Omega f^{2}rdr           \label{III:9}
\end{equation}
of an axially symmetric field configuration  of the model \eqref{II:1} in terms
of the ansatz functions.
We shall also need the expressions for  the magnetic field strength $B$ and the
radial  electric  field  strength  $E_{r}$, 
\begin{equation}
B=-\frac{a^{\prime}}{er}\quad
\text{and}\quad
E_{r}=-a_{0}^{\prime }=\frac{\Omega^{\prime }}{e}.               \label{III:10}
\end{equation}

The nontopological soliton solution of the model \eqref{II:1}  must be regular,
and its energy and electric charge must be finite.
These conditions and eqs.~\eqref{III:2}--\eqref{III:8}  lead us to the boundary
conditions for the ansatz functions:
\begin{eqnarray}
f^{\prime }\left(0\right) &=&0,\qquad f\left(\infty\right) =0,      \nonumber
 \\
a(0) &=&0,\qquad a(\infty )=a_{\infty },                            \nonumber
 \\
\Omega ^{\prime }\left( 0\right)  &=&0,\qquad \Omega (\infty )=\Omega
_{\infty },                                                      \label{III:11}
\end{eqnarray}
where  $a_{\infty}$  and  $\Omega_{\infty}$  are  constants.
Note that the function $\Omega(r) = \omega - e a_{0}(r)$ and its limiting value
$\Omega_{\infty}$  are invariant under the residual gauge transformations  with
$\Lambda = \Delta\omega t$.
Furthermore,  the  limiting  value  $\Omega_{\infty}$   is  equal  to the phase
frequency $\omega$  of  the  complex  scalar  field  $\phi$,  provided that the
gauge-fixing condition $a_{0}(\infty) = 0$ is used.

A non-topological soliton  of  the  pure  Chern-Simons gauge model \eqref{II:1}
carries a magnetic flux  $\Phi$ and  has  a  non-zero  electric  charge $Q$ and
an angular momentum $J$.
Due to the Chern-Simons  form  \eqref{III:3b} of  Gauss's  law, all these three
important  characteristics  are  expressed   in  terms  of  a  single parameter
$a_{\infty}$, the limiting  value  of  the  ansatz  function  $a(r)$ at spatial
infinity.
The corresponding formulae are
\begin{eqnarray}
\Phi  &=&-\frac{2\pi }{e}a_{\infty },                           \label{III:12a}
 \\
Q &=&-\kappa \Phi =\frac{2\pi \kappa }{e}a_{\infty },           \label{III:12b}
 \\
J &=&\frac{\pi \kappa }{e^{2}}a_{\infty }^{2} =
\frac{Q^{2}}{4\pi \kappa},                                      \label{III:12c}
\end{eqnarray}
where  eq.~\eqref{III:10}  was   used   to   obtain   eq.~\eqref{III:12a},  and
eqs.~\eqref{III:3b} and \eqref{III:9} were  used to obtain eq.~\eqref{III:12c}.
Eqs.~\eqref{III:12a} -- \eqref{III:12c} tell  us that none of $\Phi$, $Q$ , and
$J$ is quantized, since the parameter $a_{\infty}$ is not quantized in the case
of a nontopological soliton.

Eqs.~\eqref{III:12a} and \eqref{III:12b} tell us that the magnetic  flux $\Phi$
and the electric charge $Q$ of the  soliton  can change sign since the boundary
value $a_{\infty}$ can.
At the same time, $\text{sign}\left[\Phi/Q\right] = -1$, since the Chern-Simons
coupling constant $\kappa$ is assumed to be positive.
Furthermore, from eq.~\eqref{III:12c}, it follows that the angular momentum $J$
of the soliton is always positive.
These two properties are a consequence of  the fact that the model \eqref{II:1}
is not  $T$-invariant  ($P$-invariant),  since  time reversal (reflection about
an axis in the $xy$-plane) changes  both  the  sign of $\Phi/Q$ and the sign of
$J$, which contradicts   eqs.~\eqref{III:12a} -- \eqref{III:12c}.
Indeed, it is easy to show  that  the Lagrangian density in eq.~\eqref{II:1} is
not invariant  under the  $T$-transformation ($P$-transformation)  because  the
Chern-Simons  term $\mathcal{L}_{CS}   =  2^{-1}  \kappa  \epsilon^{\alpha\beta
\gamma }A_{\alpha}\partial_{\beta}A_{\gamma}$ changes sign.

We now consider the asymptotics of the soliton solution at small and large $r$.
Eqs.~\eqref{III:3b} and \eqref{III:3c}  tell  us that having a power series for
$\Omega\,$ ($a$), we automatically  obtain  a  power series for $a$ ($\Omega$).
Choosing the  ansatz   function   $\Omega$   as   an  independent  variable and
substituting  the    power   expansions   for    $f$     and    $\Omega$   into
eqs.~\eqref{III:3a} and \eqref{III:5},  we  obtain  the  asymptotics of $f$ and
$\Omega$  at small $r$,
\begin{eqnarray}
f & = &f_{0}+\frac{1}{2!}\frac{f_{0}}{2}\left[ \frac{m^{2}f_{0}^{2}}{
1+\varepsilon ^{2}}\left( 3f_{0}^{2}-4\right) +m^{2}-\Omega _{0}^{2}\right]
r^{2}+O\left[ r^{4}\right],                                     \label{III:14a}
   \\
\Omega  & = &\Omega _{0}+\frac{1}{2!}\frac{m_{A}^{2}}{2}\Omega
_{0}f_{0}^{4}r^{2}+O\left[ r^{4}\right].                        \label{III:14b}
\end{eqnarray}
Then,   substituting    eqs.~\eqref{III:14a}     and     \eqref{III:14b}   into
eq.~\eqref{III:3b} and integrating it,  we  obtain the small $r$ asymptotics of
$a$,
\begin{eqnarray}
a & = & \frac{1}{2!}m_{A}f_{0}^{2}\Omega _{0}r^{2}+\frac{3}{4!}
m_{A}f_{0}^{2}\Omega _{0}                                           \nonumber
 \\
&&\times \left[ \frac{m_{A}^{2}}{2}f_{0}^{4}+\frac{m^{2}f_{0}^{2}}{
1+\varepsilon ^{2}}\left( 3f_{0}^{2}-4\right) +m^{2}-\Omega _{0}^{2}\right]
r^{4}+O\left[ r^{6}\right].                                     \label{III:14c}
\end{eqnarray}
It follows from eqs.~\eqref{III:14a} -- \eqref{III:14c}  that  the behaviour of
the soliton solution at small $r$  is  determined by the two parameters $f_{0}$
and $\Omega_{0}$, where  $f_{0}$  can  be  considered  positive without loss of
generality.
In  particular,  eqs.~\eqref{III:14b}  and  \eqref{III:14c}  tell  us  that the
parameter $\Omega_{0}\ne 0$, since  otherwise, the ansatz functions $\Omega(r)$
and $a(r)$ vanish over the entire interval $\left[0,\infty\right)$.
We see also that the case $f_{0}= 0$ corresponds to the trivial solution: $f(r)
= 0,\,\Omega(r) = \Omega_{0}$ and $a(r) = 0$.

Linearization of eq.~\eqref{III:3a} at large $r$  and  use of the corresponding
boundary condition $f(\infty) = 0$ lead us to the large $r$ asymptotics of $f$,
\begin{equation}
f\sim c_{f}\sqrt{\frac{\pi}{2\Delta r}}e^{-\Delta r}\left(1+O\left[\left(\Delta
r\right)^{-1}\right] \right),                                    \label{III:15}
\end{equation}
where $\Delta = \left(m^{2} - \Omega_{\infty}^{2}\right)^{1/2}$.
From eq.~\eqref{III:15}  and  the  definition  of  the  parameter  $\Delta$, it
follows that $\left\vert \Omega_{\infty}\right\vert\le m$, since otherwise, the
solution would be oscillatory.
Substituting  eq.~\eqref{III:15}  into  eqs.~\eqref{III:5}  and  \eqref{III:6},
retaining the leading terms, and  solving  the  resulting  linear  differential
equations, we obtain the large $r$ asymptotics of $\Omega$ and $a$,
\begin{eqnarray}
\Omega  &\sim &\Omega _{\infty }+\frac{c_{\Omega }}{r}\frac{e^{-2\Delta r}}{
2\Delta r}\left( 1+O\left[ \left( \Delta r\right) ^{-1}\right] \right),
                                                                 \label{III:16}
  \\
a &\sim &a_{\infty }+c_{a}e^{-2\Delta r}\left( 1+O\left[ \left( \Delta
r\right) ^{-1}\right] \right).                                   \label{III:17}
\end{eqnarray}
Then, using   eqs.~\eqref{III:3b}   and   \eqref{III:3c},  we   find   that  in
eq.~\eqref{III:17}, the  asymptotic  parameters $a_{\infty}$ and $c_{a}$ can be
expressed  in   terms   of   $c_{f}$,  $c_{\Omega}$,  and  $\Omega_{\infty}$ as
\begin{equation}
a_{\infty }=-\frac{2}{\pi }\frac{\Delta }{m_{A}}\frac{c_{\Omega }}{c_{f}^{2}}
\quad \text{and} \quad
c_{a}=-\frac{\pi }{4}\frac{m_{A}}{\Delta ^{2}}c_{f}^{2}\Omega _{\infty}.
                                                                 \label{III:18}
\end{equation}

From the above, it follows  that  for a fixed value of $\Omega_{\infty}$ in the
boundary conditions \eqref{III:11},  the  large  $r$ asymptotics of the soliton
solution  is  determined  by   the   two  independent  parameters,  $c_{f}$ and
$c_{\Omega}$.
At the same time, the small $r$ asymptotics  of  the  soliton  solution is also
determined by the two parameters, $f_{0}$ and $\Omega_{0}$.
To determine the four parameters $c_{f}$, $c_{\Omega}$, $f_{0}$, and $\Omega_{0
}$, we impose four independent  matching conditions, namely the equality of the
matched ansatz functions  and  their  derivatives at an arbitrary radial point.
The fact that the number of  parameters coincides with the number of conditions
determining them is   an   argument  in  favour of the existence of the soliton
solution.
Eqs.~\eqref{III:15} -- \eqref{III:17}  show  that  the exponential decay of the
ansatz functions is determined solely by  the parameter $\Delta = \left(m^{2} -
\Omega_{\infty}^{2}\right)^{1/2}$.
Hence, the large $r$ asymptotics  of  the  soliton's  gauge  field  is governed
by the mass $m$ of the scalar field in the unbroken vacuum.
The reason is that there are  no propagating gauge modes in the unbroken vacuum
of the model \eqref{II:1}, whereas there are two propagating scalar modes.

Eqs.~\eqref{III:3b} -- \eqref{III:3c}  and   \eqref{III:14b} -- \eqref{III:14c}
allow us to draw a number of  conclusions  about  the  global  behaviour of the
ansatz functions $\Omega(r)$ and $a(r)$.
Indeed, from eqs.~\eqref{III:14b} -- \eqref{III:14c},  it follows that at least
for sufficiently small $\delta$,
\begin{equation}
\text{sign}\left[ \Omega ^{\prime }\left( r\right) \right] =
\text{sign}\left[a^{\prime }\left( r\right) \right] =
\text{sign}\left[ \Omega \left( r\right) \right] =
\text{sign}\left[ a\left( r\right) \right] =
\text{sign}\left[ \Omega_{0} \right]                             \label{III:19}
\end{equation}
for $r \in (0, \delta)$.
At the same time,  eqs.~\eqref{III:3b} -- \eqref{III:3c}  tell  us that for all
$r$ such that $f(r) \ne 0$,
\begin{equation}
\text{sign}\left[ a^{\prime }\left( r\right) \right]=\text{sign}\left[
\Omega \left( r\right) \right] \quad \text{and}\quad \text{sign}\left[
\Omega ^{\prime }\left( r\right) \right]=\text{sign}\left[a\left(r\right)
\right].                                                         \label{III:20}
\end{equation}
Let us suppose that $f(r)$ vanishes at $r = \bar{r}$.
Then it follows from eqs.~\eqref{III:3b} -- \eqref{III:3c} that $\Omega^{\prime
}\left( \bar{r}\right) = a^{\prime }\left( \bar{r}\right) = 0$.
Moreover,  differentiating  eqs.~\eqref{III:3b} -- \eqref{III:3c}  successively
with respect to $r$, we find  that $\Omega^{\prime \prime}\left( r \right)$ and
$a^{\prime \prime}\left( r \right)$ also vanishes at $r = \bar{r}$, whereas the
higher derivatives of  $\Omega(r)$  and  $a(r)$ are different from zero at this
point.
It follows that the point $r = \bar{r}$ can be neither a minimum nor a maximum,
but only an extremum point of $\Omega(r)$ and  $a(r)$.
Combining this with  eqs.~\eqref{III:19}  and  \eqref{III:20}, we arrive at the
conclusion that both $\Omega(r)$ and $a(r)$ increase  (decrease)  monotonically
over  the   entire  interval  $[0,  \infty)$,  provided  that  $\Omega_{0} > 0$
($\Omega_{0} < 0$).
Hence,  $\Omega(r)$   does   not   vanish   on   the  interval  $[0,  \infty)$.

The soliton’s energy \eqref{III:8} can  be  written  as the sum of three terms,
\begin{equation}
E = E_{T} + E_{G} + E_{P},                                       \label{III:21}
\end{equation}
where
\begin{equation}
E_{T}=2\pi v^{2}\int_{0}^{\infty } \Omega ^{2}f^{2}rdr           \label{III:22}
\end{equation}
is the kinetic part of the energy,
\begin{equation}
E_{G}=2\pi v^{2}\int_{0}^{\infty }
       \left[ f^{\prime 2}+\frac{a^{2}}{r^{2}}f^{2}\right] rdr   \label{III:23}
\end{equation}
is the gradient part of the energy, and
\begin{equation}
E_{P}=2\pi v^{2}\int_{0}^{\infty}\tilde{V}\left( f\right) rdr    \label{III:24}
\end{equation}
is the potential part of the energy.
The Lagrangian $L = \int \mathcal{L}d^{2}x$ can  also  be expressed as a linear
combination of these three terms plus the Chern-Simons term,
\begin{equation}
L=L_{CS}+E_{T}-E_{G}-E_{P},                                      \label{III:25}
\end{equation}
where
\begin{equation}
L_{CS}=\frac{\pi \kappa }{e}\int_{0}^{\infty }\left[ a_{0}a^{\prime
}-a_{0}^{\prime }a\right] dr.                                    \label{III:26}
\end{equation}
Integrating by  parts  and  using  Gauss's  law  \eqref{III:3b},  we  write the
Chern-Simons term  as $L_{CS}=\Omega _{\infty }Q_{N}-2E_{T}$, where the Noether
charge $Q_{N} = Q/e$.
Using this form  of  $L_{CS}$,  we  rewrite  the  Lagrangian  \eqref{III:25} as
\begin{equation}
L = \Omega_{\infty}Q_{N} - E.                                    \label{III:27}
\end{equation}

By definition, the  soliton  solution  is  an  extremum of the action $S = \int
\mathcal{L}d^{2}xdt$.
From eqs.~\eqref{III:22} -- \eqref{III:27},  it  follows  that  the  Lagrangian
density  corresponding  to  the  soliton  solution  is  time-independent, which
implies that $S \propto L$.
Hence, the soliton solution  is  also  an  extremum of the Lagrangian $L = \int
\mathcal{L}d^{2}x$.
Combining this fact and  eq.~\eqref{III:27},  we come to the conclusion that in
the functional neighbourhood of the soliton solution, the following variational
relation holds
\begin{equation}
\delta L=\Omega _{\infty }\delta Q_{N}-\delta E.                 \label{III:28}
\end{equation}
From eq.~\eqref{III:28},  it  follows  that  the  the  soliton  solution  is an
extremum of the energy functional $E$, provided that the Noether charge $Q_{N}$
is fixed.
We see also  that  the   parameter  $\Omega_{\infty}$,  the  boundary  value of
$\Omega(r)$  at  spatial  infinity,  plays  the  role of a Lagrange multiplier.
According to the method of Lagrange  multipliers,  the variations of the fields
in eq.~\eqref{III:28} are not restricted by the condition $Q_{N}=\text{const}$.
In particular, these  variations  can  link  two  infinitesimally close soliton
solutions.
In this  case,  Eq.~\eqref{III:28}   results   in   the  differential  relation
\begin{equation}
dE/dQ_{N} = \Omega _{\infty}.                                    \label{III:29}
\end{equation}
Eq.~\eqref{III:29} relates the  derivative of the soliton’s energy with respect
to the Noether charge with the  boundary  value of  the function $\Omega(r)$ at
spatial infinity.
Similar relations  are  also  valid  for  nontopological solitons in many other
models.
Combining eq.~\eqref{III:29} with  eqs.~\eqref{III:12b} and \eqref{III:12c}, we
obtain a few more differential relations,
\begin{eqnarray}
dE/d\Phi  &=&-\kappa e^{-1}\Omega _{\infty },                   \label{III:30a}
 \\
dE/dJ &=&2\pi \kappa e^{-2}\Omega_{\infty}/Q_{N},               \label{III:30b}
 \\
dE/da_{\infty} &=& 2\pi \kappa e^{-2}\Omega_{\infty}.           \label{III:30c}
\end{eqnarray}

In addition to eqs.~\eqref{III:21} and \eqref{III:25}, the energy parts $E_{T}$
and $E_{P}$ are related by one more linear (virial) relation,
\begin{equation}
E_{T}  - E_{P } = 0.                                             \label{III:31}
\end{equation}
To derive eq.~\eqref{III:31}, we perform the scale transformation $r\rightarrow
\varkappa r$ of  the  argument  of  the  ansatz  functions  $f(r)$, $a(r)$, and
$a_{0}(r)$ of the soliton solution.
After this, the  Lagrangian  $L$  becomes  a  function  of  the scale parameter
$\varkappa$.
The function $L(\varkappa)$ has a stationary point  at $\varkappa=1$, since the
soliton solution is an unconditional extremum of the Lagrangian.
Hence,        $\left.  dL/d\varkappa\right\vert_{\varkappa = 1}      =      0$.
Furthermore, it can easily be shown that $E_{T}\rightarrow\varkappa^{-2}E_{T}$,
$E_{G}  \rightarrow E_{G}$, $E_{P}\rightarrow\varkappa ^{-2}E_{P}$, and $L_{CS}
\rightarrow L_{CS}$  under  the  rescaling $r \rightarrow \varkappa r$.
Using  these    scale   transformation   rules,   eq.~\eqref{III:25},   and the
stationarity condition, we obtain the virial relation \eqref{III:31}.

It was shown  in  \cite{hong_prl_1990, jw_prl_1990}  that in the self-dual case
of $m_{A} = m_{H} = 2 m$, the  ansatz  functions  $a(r)$ and $f(r)$ satisfy the
system of the first-order equations
\begin{eqnarray}
\hat{f}^{\prime } &=&\pm \frac{\hat{a}\hat{f}}{r},              \label{III:32a}
\\
\hat{a}^{\prime } &=&\pm 2m^{2}r\hat{f}^{2}\bigl( \hat{f}
^{2}-1\bigr),                                                   \label{III:32b}
\end{eqnarray}
where the upper (lower)  sign  corresponds  to  a  positive (negative) value of
$\Phi$ and we use the hat notation to distinguish the self-dual  case  from the
general non-self-dual case.
The self-dual   ansatz   functions   also   satisfy   the   second-order system
\eqref{III:3a} -- \eqref{III:3c}.
In particular, we can use Gauss's law \eqref{III:3b} and eq.~\eqref{III:32b} to
eliminate $\hat{a}^{\prime}$, resulting in the equation
\begin{equation}
\hat{\Omega }=\mp m \bigl( 1-\hat{f}^{2}\bigr).                  \label{III:33}
\end{equation}
Eq.~\eqref {III:33} and  the  boundary  condition $\hat{f}(\infty) = 0$ tell us
that the boundary value $\hat{\Omega}_{\infty} = \mp m$.
Then, it  follows  from  eqs.~\eqref{III:29}  that  in  the self-dual case, the
energy and  the  Noether  charge  of  the  nontopological  soliton are linearly
related, $E = m \vert Q_{N} \vert$.

\section{Numerical results}                                       \label{sec:V}

We have established above that  the system of three equations \eqref{III:3a} --
\eqref{III:3c}  is  equivalent  either   to   the  system  of  two second-order
equations \eqref{III:3a} and \eqref{III:5} or to the system of two second-order
equations \eqref{III:3a} and \eqref{III:6}.
To study the properties  of   the   nontopological  soliton,  we choose the two
second-order differential  equations \eqref{III:3a} and \eqref{III:5}  with the
corresponding boundary conditions from eq.~\eqref{III:11}.
Additionally, the  term  $a^{2} f/r^{2}$  in eq.~\eqref{III:3a} is expressed in
terms of $\Omega^{\prime}$ according to eq.~\eqref{III:3c}.
As a result, we obtain  a  mixed  boundary  value problem  on the semi-infinite
interval $[0, \infty)$, which  can  be  solved  only  using  numerical methods.
We used  the   numerical   methods   realized   in   the  {\sc{Maple}}  package
\cite{maple_2022} to solve this bvp problem.
To verify  the  correctness   of   the   numerical  solutions,  we  checked the
validity of eqs.~\eqref{III:29} and \eqref{III:31}.

From eqs. \eqref{III:3a}, \eqref{III:5},  and  \eqref{III:11},  it follows that
the bvp problem depends on four parameters three of which , $m$, $m_{A}=2 e^{2}
v^{2}\kappa^{-1}$, and $\Omega_{\infty}$ ,  have  the  dimension of mass, while
the fourth, $\varepsilon$, is dimensionless.
Let us  introduce the two dimensionless   parameters  $\tau  =  m_{A}/(2m)$ and
$\bar{\Omega}_{\infty}  =  \Omega_{\infty}/m$,  as  well  as  the dimensionless
radial variable $\rho = m r$.
Then the two dimensionless ansatz functions can be written as $f\left(r\right)=
\bar{f}\left( \rho,  \varepsilon,  \tau,  \bar{\Omega }_{\infty }  \right)$ and
$a\left(r\right) =\bar{a}\left( \rho, \varepsilon, \tau, \bar{\Omega }_{\infty}
\right)$, while the function  $\Omega(r)$, which has the dimension of mass, can
be written as $\Omega\left(r\right)=m \bar{\Omega }\left(\rho,\varepsilon,\tau,
\bar{\Omega}_{\infty } \right)$,   where   $\bar{\Omega}$   is   dimensionless.
Using these   representations    of    the   ansatz   functions   together with
eqs.~\eqref{III:7} and \eqref{III:8}, we  find that the Noether charge  and the
energy of the soliton can be written as
\begin{equation}
Q_{N}=v^{2}m^{-1}\bar{Q}_{N}\left(\varepsilon,\tau,\bar{\Omega}_{\infty}\right)
\quad\text{and}\quad E=v^{2}\bar{E}\left(\varepsilon,\tau,\bar{\Omega}_{\infty}
\right),                                                            \label{V:1}
\end{equation}
respectively,   where    $\bar{Q}_{N}$   and   $\bar{E}$   are   dimensionless.
In this paper, the results of the numerical calculations are presented in terms
of dimensionless quantities.

From  the     results      of       section~\ref{sec:III},   it   follows  that
$\bar{\Omega}\left(\rho\right)\gtrless 0$ and $\bar{a}\left(\rho\right)\gtrless
0$, provided that $\bar{\Omega}_{\infty}\gtrless 0$.
It follows that the  signs  of  the boundary values $\bar{\Omega}_{\infty}$ and
$\bar{a}_{\infty} \equiv a_{\infty}$ coincide.
Furthermore it follows  from the above and  eq.~\eqref{III:7}  that the Noether
charge $\bar{Q}_{N}\gtrless 0$, provided that $\bar{\Omega}_{\infty}\gtrless 0$.
In addition,  eq.~\eqref{II:2}  tells  us  that  the  sign  of $\varepsilon$ is
irrelevant.
In what  follows,   we   assume  that  the  Noether  charge  $\bar{Q}_{N}$, the
boundary values $\bar{\Omega}_{\infty}$  and  $a_{\infty}$,  and  the parameter
$\varepsilon$ are positive.
To apply numerical results obtained  to  the  case  of  negative $\bar{Q}_{N}$,
$\bar{\Omega}_{\infty}$, and $a_{\infty}$,  we  must  replace  them  with their
absolute values.

\begin{figure}[tbp]
\begin{center}
\includegraphics[width=0.6\textwidth]{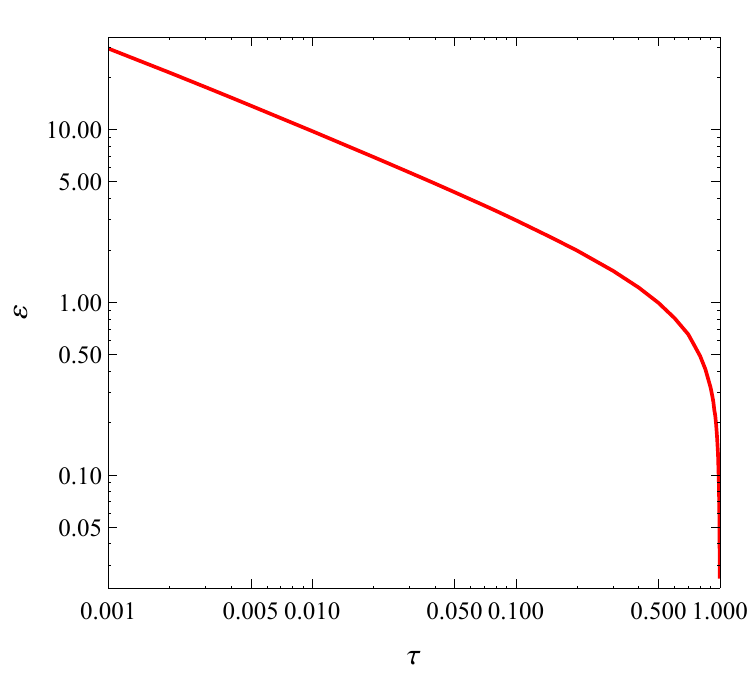}
\caption{\label{fig1}    The existence  region  of  the soliton solution in the
$\tau\varepsilon$ plane.    The function $\varphi_{\text{b}}(\tau) $ (red solid
curve) bounds the existence region from above.}
\includegraphics[width=0.6\textwidth]{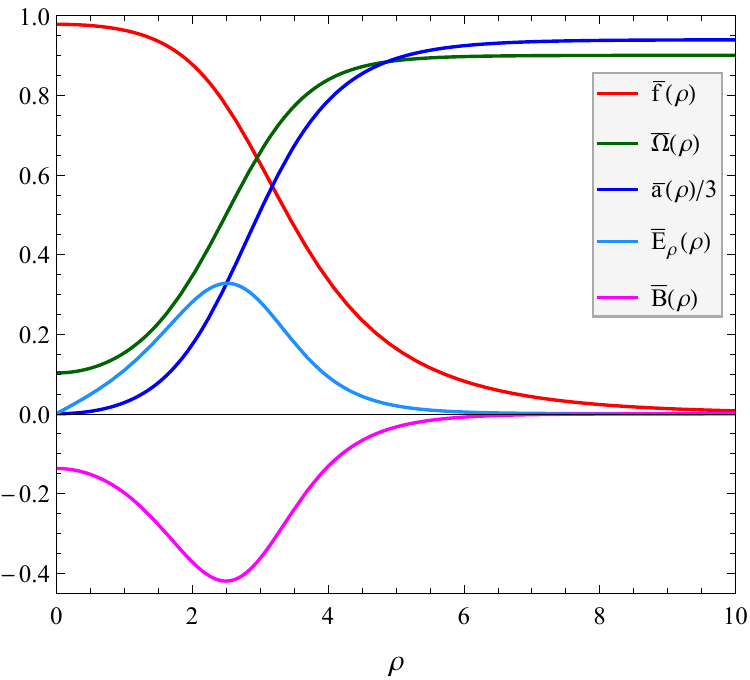}
\caption{\label{fig2}    The dimensionless  ansatz  functions  $\bar{f}(\rho)$,
$\bar{\Omega}(\rho)$,  $\bar{a}(\rho)/3$,   the  radial electric field strength
$\bar{E}_{\rho}(\rho)$, and  the  magnetic field  strength $\bar{B}(\rho)$ of a
nontopological soliton solution.     The parameters of the soliton solution are
$\varepsilon = 0.2$, $\tau = 0.7$, and $\bar{\Omega}_{\infty} = 0.9$.}
\end{center}
\end{figure}

We first discuss  the  region  of  the  parameters $\tau$ and $\varepsilon$, in
which the nontopological soliton can exist.
The result of the numerical studies of this issue is shown in figure~\ref{fig1},
where  both  the  vertical  and  horizontal  axes  are  scaled  logarithmically
for readability.
We see that the  domain  of  existence  of  the  soliton  solution is the first
quadrant  of  the   $\tau  \varepsilon$-plane,   bounded  above  by  a function
$\varepsilon = \varphi_{\text{b}}(\tau)$.
The  boundary   function  $\varepsilon  =  \varphi_{\text{b}}(\tau)$  decreases
monotonically on the interval $(0, 1]$.
It increases  unboundedly  $\propto  \tau^{-1/2}$  as $\tau \rightarrow 0$, and
tends to zero $\propto (1-\tau)^{1/2}$ as $\tau \rightarrow 1$.
Recall that the limiting point $(\tau, \varepsilon) = (1,0)$ corresponds to the
self-dual   case    considered    in   refs.~\cite{hong_prl_1990,  jw_prl_1990,
 jlw_prd_1990}.
The opposite  case  of  the  vanishing   parameter  $\tau$  corresponds  to the
vanishing gauge coupling constant $e$.
In this case, the gauge potential $A_{\mu}$  decouples  from the scalar matter,
becoming  a  pure  gauge  field  configuration  with  the  zero  field strength
$F_{\mu \nu}$.
For this  reason,   the   nontopological   soliton   turns   into   an ungauged
two-dimensional  Q-ball,   which  exists  for  all  $\varepsilon$  lying in the
interval $[0, \infty)$.

It is worth noting that we  were  unable  to find the soliton solution when the
parameter $\tau >1$,  and  therefore  came  to  the  conclusion that no soliton
solution exists in this region.
This is consistent with  the  fact  that  the  boundary function $\varepsilon =
\varphi_{\text{b}}(\tau)$ has an  algebraic  branch  point at $\tau = 1$, since
$\varphi_{\text{b}}(\tau) \propto (1 - \tau)^{1/2}$  as  $\tau  \rightarrow 1$.
It follows  that  the function  $\varphi_{\text{b}}(\tau)$  must  have a branch
cut $[1,\infty)$ in the complex $\tau$ plane.
In the  general  case,  the   function  $\varphi_{\text{b}}(\tau)$  acquires an
imaginary part when analytically continued to the upper  (lower)  side  of  the
branch cut $[1, \infty)$.
The imaginary part of the  boundary  function  $\varphi_{\text{b}}(\tau)$ makes
the existence   of  the  soliton   solution   impossible,  since  the parameter
$\varepsilon$ must be real.

At every admissible point $(\tau,\varepsilon)$ of the $\tau \varepsilon$-plane,
the soliton  solution  exists  in  the  finite  interval $(\bar{\Omega}_{\infty
}^{\min}, 1]$ of $\bar{\Omega}_{\infty}$, where the  minimum value $\bar{\Omega
}_{\infty}^{\min}$  is a function of $\tau$ and $\varepsilon$.
In particular, the minimum value $\bar{\Omega}_{\infty}^{\min}$  tends to unity
as the  point  $(\tau, \varepsilon)$   tends   to   a   point  on  the boundary
$\varepsilon = \varphi_{\text{b}}(\tau)$.
As a result, the interval $(\bar{\Omega}_{\infty}^{\min},1]$ shrinks to a point
and the soliton solution ceases to exist.

The self-dual point $(\tau,\varepsilon) = (1,0)$ must be considered separately.
This case was studied in refs.~\cite{hong_prl_1990, jw_prl_1990, jlw_prd_1990},
where it was shown  that  at  the  self-dual  point,  the  energy  of a soliton
carrying  the  magnetic flux $\Phi$  reaches  the minimum  possible  value $E =
e v^{2}\left\vert \Phi \right\vert  =  e v^{2}\kappa^{-1} Q  =  m_{A} Q_{N}/2$.
The last expression can be rewritten as $E = m Q_{N}$, since $\tau \equiv m_{A
}/\left(2m\right) = 1$ at the self-dual point.
Then, it  follows from eq.~\eqref{III:29}  that  at  the  self-dual  point, the
boundary value $\Omega _{\infty } = dE/d Q_{N} = m$, which is equivalent to the
dimensionless relation $\bar{\Omega}_{\infty } = 1$.
This result  is  consistent with eq.~\eqref{III:33}.
Unlike $\Omega_{\infty}$, the boundary  value $a_{\infty}$ is not fixed for the
self-dual nontopological soliton.
Indeed, it was shown  in  \cite{jlw_prd_1990}  that  in this case, the boundary
value $a_{\infty} \in (2, \infty)$.
Thus, despite the fact that the interval of $\Omega_{\infty}$ has shrunk to the
point in  the  self-dual  case,  there  exists  a  family  of soliton solutions
parameterized by the boundary value $a_{\infty}$.
Note that  in  the  non-self-dual  case, the parameters $\bar{\Omega}_{\infty}$
and $a_{\infty}$ are related by a functional dependence, $\bar{\Omega}_{\infty}
= h\left( a_{\infty }\right)$. 
In contrast,  in  the  self-dual  case  of  $(\tau,\varepsilon)  =  (1,0)$, the
parameter $\bar{\Omega}_{\infty}$  ceases  to  depend  on $a_{\infty}$, and the
function $h\left(a_{\infty}\right)$ degenerates into the constant equal to $1$.

We define   the  dimensionless   electric   and   magnetic  field  strengths as
$\bar{\mathbf{E}} = e m^{-2}\mathbf{E}$ and $\bar{B} = em^{-2}B$, respectively.
Figire~\ref{fig2} presents the dimensionless ansatz functions  $\bar{f}(\rho)$,
$\bar{\Omega}(\rho)$, $\bar{a}(\rho)/3$ (scaled  for  readability),  the radial
electric field strength $\bar{E}_{\rho}(\rho)$, and the magnetic field strength
$\bar{B}(\rho)$ of a nontopological soliton solution.
The  soliton   solution   corresponds   to   the non-extreme  parameter  values
$\varepsilon = 0.2$, $\tau = 0.7$, and $\bar{\Omega}_{\infty} = 0.9$.
We see that the behaviour  of  the  ansatz  functions  in  figure~\ref{fig2} is
consistent with the conclusions of section~\ref{sec:III}.
In particular,   both   $\bar{\Omega}(\rho)$   and   $\bar{a}(\rho)$   increase
monotonically, and the behaviour  of  the  ansatz  functions at small and large
$\rho$ is consistent with eqs.~\eqref{III:14a} -- \eqref{III:17}.
Figure~\ref{fig2} shows that the magnetic  field  of the nontopological soliton
reaches a maximum at a finite $\rho$, which is a characteristic property of the
Chern-Simons gauge models \cite{dunne_1995, horvathy_pr_2009}.
The radial electric field  also  has  a  maximum  at  a  finite  $\rho$, and it
follows from eq.~\eqref{III:10} that the position of the maximum coincides with
that of the inflection point of $\bar{\Omega}(\rho)$.
Unlike the magnetic field,  the  radial  electric field vanishes at the origin,
which is a consequence of the regularity of the soliton solution.

\subsection{The case of nonzero $\varepsilon$}

\begin{figure}[tbp]
\begin{center}
\includegraphics[width=0.6\textwidth]{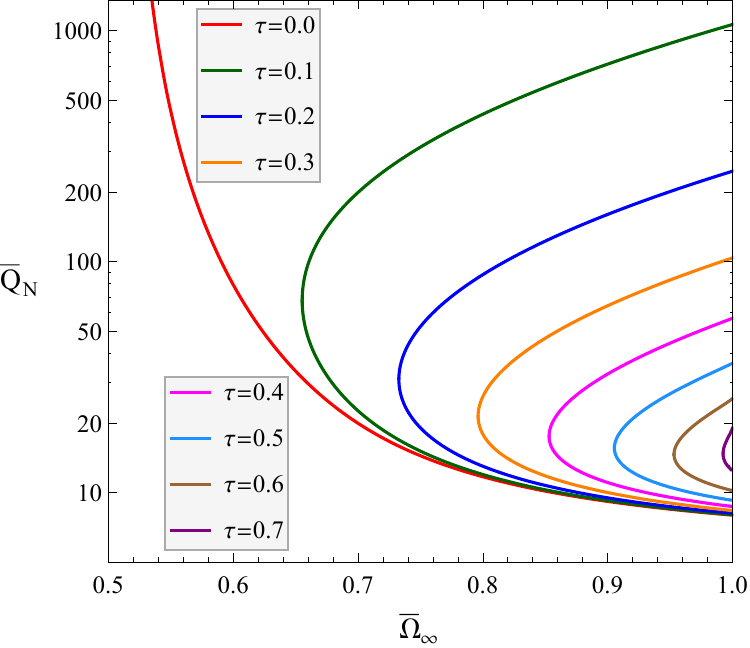}
\caption{\label{fig3}   Dependencies of the Noether charge $\bar{Q}_{N}$ of the
nontopological  soliton   on    $\bar{\Omega}_{\infty}$   corresponding  to the
parameter $\varepsilon = 0.6$ and different  values  of  the  parameter $\tau =
e^{2}v^{2}/(\kappa m)$.}
\includegraphics[width=0.6\textwidth]{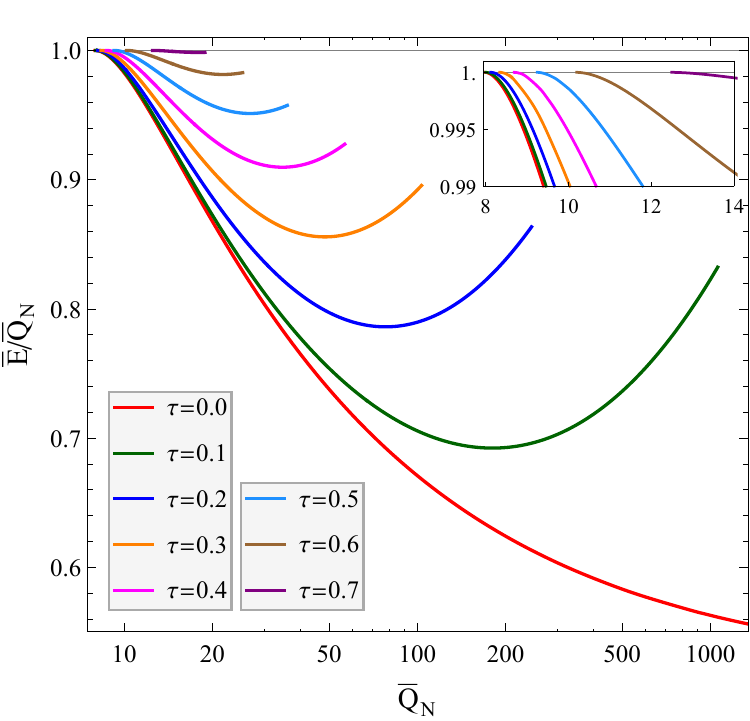}
\caption{\label{fig4}  Dependencies of the ratio of the energy $\bar{E}$ of the
nontopological soliton  to  its  Noether charge $\bar{Q}_{N}$ on $\bar{Q}_{N}$.
The values of the parameters $\varepsilon$ and $\tau$ are the same as in figure
\ref{fig3}.}
\end{center}
\end{figure}

We  now    turn     to     the     study    of    the    soliton's  properties.
We shall  first  consider  the  case   of  nonzero  $\varepsilon$  in which the
potential \eqref{II:2} has a single zero global minimum at $\phi= 0$, and hence
we are above a first-order phase transition point.
Figure~\ref{fig3} shows the  dependencies  of  the Noether charge $\bar{Q}_{N}$
of the nontopological soliton on  the  boundary  value $\bar{\Omega}_{\infty}$,
which coincides with the phase frequency $\omega$ of the scalar field $\phi$ in
the used gauge $a_{0}(\infty) = 0$.
The curves   $\bar{Q}_{N}(\bar{\Omega}_{\infty})$  correspond  to the parameter
$\varepsilon = 0.6$ and  different  values  of  the parameter $\tau = e^{2}v^{2
}/(\kappa m)$.
In particular, the case of $\tau = 0$  corresponds  to  the  decoupling  of the
Chern-Simons gauge field from the matter (complex scalar) field, so that we are
dealing with an usual nongauged two-dimensional Q-ball.
In this  trivial  case,  the  Noether  charge  $\bar{Q}_{N}$ grows  unboundedly
$\propto \left(\bar{\Omega}_{\infty} -\bar{\Omega}_{\infty}^{\min}\right)^{-2}$
as $\bar{\Omega}_{\infty}\rightarrow\bar{\Omega}_{\infty}^{\min}$, where $\bar{
\Omega}_{\infty}^{\min}  =  \varepsilon \left(1 + \varepsilon^{2}\right)^{-1/2}
\approx 0.5145$.
Note that the above expression for $\bar{\Omega}^{\min}_{\infty}$ is valid only
for the nongauged case of $\tau = 0$.
When $\tau > 0$, and hence the gauge interaction is turned on, the behaviour of
the curves $\bar{Q}_{N}(\bar{\Omega}_{\infty})$ changes drastically.
Specifically, the Noether charge of any gauged  Q-ball  remains  finite for all
$\bar{\Omega}_{\infty}\in \left[\bar{\Omega}_{\infty }^{\min}, 1\right]$, where
the  minimum   possible   value   $\bar{\Omega}_{\infty }^{\min}$   depends  on
$\varepsilon$ and $\tau$.
Furthermore, we see that  for  a given $\bar{\Omega}_{\infty}$, there exist the
two gauged Q-ball solutions whose Noether charges  and  energies  can differ by
several orders of magnitude.
In particular, the  curves  $\bar{Q}_{N}(\bar{\Omega}_{\infty})$  intersect the
limiting line  $\bar{\Omega}_{\infty}  =  1$  at  two  points,  where the upper
intersection  point  (the  Noether  charge $\bar{Q}^{\text{up}}_{N}(1)$  of the
corresponding soliton) tends to infinity as $\tau \rightarrow 0$.
Specifically, it was found numerically that $\bar{Q}^{\text{up}}_{N}(1) \propto
\tau^{-2}$ as $\tau \rightarrow 0$.
With an increase in $\tau$, $\bar{\Omega}_{\infty }^{\min}$ also increases, and
tends  to  unity   as $\tau  \rightarrow  \varphi_{\text{b}}^{-1}(\varepsilon)$.
In accordance with figure~\ref{fig1}, there is no soliton solution if $\tau \ge
\varphi_{\text{b}}^{-1}(\varepsilon)$.

Except  for the curve  corresponding to the nongauged  case $\tau = 0$, all the
curves in figure~\ref{fig3} have  turning   points,  at  which   the derivative
$d\bar{Q}_{N}/d\bar{\Omega}_{\infty}$ becomes infinite.
Hence, the derivative of  the  inverse function $d\bar{\Omega}_{\infty}/d\bar{Q
}_{N}$ vanishes at these points.
Eq.~\eqref{III:29} tells us that the derivative $d\bar{\Omega}_{\infty}/d\bar{Q
}_{N} = d^{2}\bar{E}/d\bar{Q}^{2}_{N}$, and  hence the second derivative $d^{2}
\bar{E}/d\bar{Q}^{2}_{N}$ vanishes at the turning points.
It follows that  the  turning  points  of  the  curves $\bar{Q}_{N}(\bar{\Omega
}_{\infty})$ correspond to the inflection  points  of  the corresponding curves
$\bar{E}(\bar{Q}_{N})$.
The curve  $\bar{E}(\bar{Q}_{N})$ therefore  has  one  inflection  point in the
gauged  case,  whereas  it  has  no  inflection  points  in the nongauged case.
Note that the behaviour of  the  curves in figure~\ref{fig3} is similar to that
of the corresponding  curves  for  an  electrically  charged  three-dimensional
Q-ball \cite{gulamov_prd_2015, loginov_prd_2020}.

Figure~\ref{fig4} presents  the  dependence  of the ratio $\bar{E}/\bar{Q}_{N}$
on the Noether  charge  $\bar{Q}_{N}$  for  $\varepsilon = 0.6$  and  different
values of $\tau$.
We see that  as  in  figure~\ref{fig3}, the  curve  for  the nongauged $Q$-ball
differs significantly  from  those  for  the  gauged  nontopological  solitons.
In particular, the  energy and Noether  charge  of   the   $Q$-ball
can take arbitrarily large values, while there are upper bounds on  the  energy
and  Noether  charge of the nontopological  solitons.
For   each    of    the    curves   shown   in   figure~\ref{fig3},  the  ratio
$\bar{E}/\bar{Q}_{N}$  is  less than unity, and it tends to unity from below as
$\bar{Q}_{N}$ tends to the minimum value $\bar{Q}_{N}^{\min}$.
It follows that both nongauged  $Q$-ball  and  gauged  solitons are stable with
respect to the transition into  a  configuration of the scalar $\phi$-particles
of mass $m$, since the minimum  possible  energy of this configuration is equal
to $\bar{Q}_{N}$ in the dimensionless notations used.
We found numerically that for each curve in figure~\ref{fig3},  the combination
$\bar{E}/\bar{Q}_{N} - 1 \propto - (\bar{Q}_{N}  -  \bar{Q}_{N}^{\min})^{2}$ as
$\bar{Q}_{N} \rightarrow \bar{Q}_{N}^{\min}$.

In figure~\ref{fig4}, each of the curves corresponding to gauged solitons has a
global minimum (visually indistinguishable for $\tau = 0.7$), whereas the curve
that corresponds to the nongauged $Q$-ball has no minimum.
This is consistent with the fact that in figure~\ref{fig3}, all the curves have
turning poins, except for  the  curve  corresponding  to  the nongauged case of
$\tau = 0$.
Indeed, it is easy to show that at the minimum point of a curve $\bar{E}(\bar{Q
}_{N})/\bar{Q}_{N}$,   the  derivative  $d\bar{E}/d\bar{Q}_{N} = \bar{E}/\bar{Q
}_{N}$.
It follows that the  minimum  point of  the  curve $\bar{E}(\bar{Q}_{N})/\bar{Q
}_{N}$ corresponds to the  point  of  contact between the curve $\bar{E}(\bar{Q
}_{N})$ and  the straight line passing through the origin of the $\bar{E}\bar{Q
}_{N}$-plane.
It is obvious that this contact  point  must lie to the right of the inflection
point on the curve $\bar{E}(\bar{Q}_{N})$.
Hence, the existence of an inflection point on the curve $\bar{E}(\bar{Q}_{N})$
is a necessary  condition  for  the  existence   of   a  minimum  point  on the
corresponding curve $\bar{E}(\bar{Q}_{N})/\bar{Q}_{N}$.
In turn, it was shown above  that  the  presence of the inflection point on the
curve  $\bar{E}(\bar{Q}_{N})$   was  equivalent  to the presence of the turning
point on the curve $\bar{Q}_{N}(\bar{\Omega}_{\infty})$.
Thus, the  absence  of  a  turning   point   on   the  nongauged (red) curve in
figure~\ref{fig3} is the  reason  for  the  absence  of  a minimum point on the
corresponding curve in figure~\ref{fig4}.
Instead,  the  nongauged  curve  $\bar{E}(\bar{Q}_{N})/\bar{Q}_{N}$  tends from
above toward some limit value as $\bar{Q}_{N} \rightarrow \infty$.
Using eq.~\eqref{III:29},  we  conclude  that  this limit value is $\bar{\Omega
}_{\infty}^{\min} = \varepsilon \left(1 + \varepsilon^{2}\right)^{-1/2} \approx
0.5145$.

\begin{figure}[tbp]
\begin{center}
\includegraphics[width=0.6\textwidth]{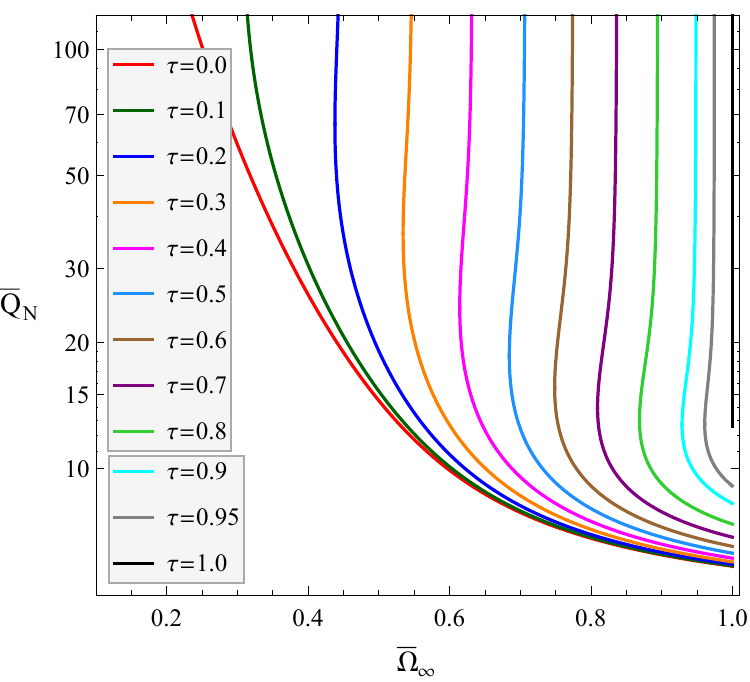}
\caption{\label{fig5}   Dependencies of the Noether charge $\bar{Q}_{N}$ of the
nontopological   soliton   on   $\bar{\Omega}_{\infty}$   corresponding  to the
parameter $\varepsilon = 0$  and  different  values  of  the  parameter $\tau =
e^{2}v^{2}/(\kappa m)$.}
\includegraphics[width=0.6\textwidth]{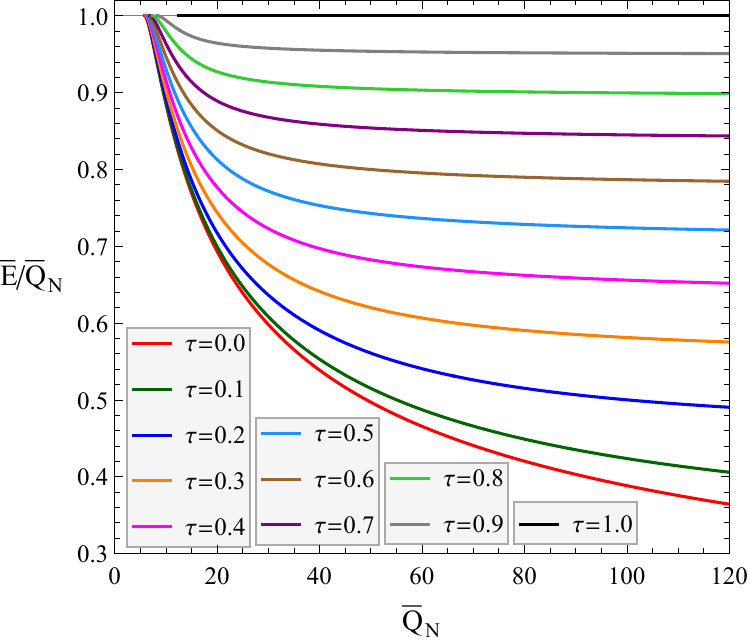}
\caption{\label{fig6}  Dependencies of the ratio of the energy $\bar{E}$ of the
nontopological soliton  to  its  Noether charge $\bar{Q}_{N}$ on $\bar{Q}_{N}$.
The values of the parameters $\varepsilon$ and $\tau$ are the same as in figure
\ref{fig5}.}
\end{center}
\end{figure}


Let us define the binding energy of a soliton as $\bar{E}_{\text{b}}=\bar{Q}_{N
} - \bar{E}$,  then   the   specific   binding   energy   of   the   soliton is
$\bar{E}_{\text{b}}/\bar{Q}_{N} = 1 - \bar{E}/\bar{Q}_{N}$.
From figure~\ref{fig4},  it  follows  that  for  a  given  $\tau$, the specific
binding  energy   is   maximal   at  the  minimum  point  of  the corresponding
$\bar{E}(\bar{Q}_{N})/\bar{Q}_{N}$ curve.
With further growth of $\bar{Q}_{N}$,  the  specific  binding energy decreases,
reaching a local minimum at the right boundary point of the curve.
Consider a soliton of charge $\bar{Q}_{N}$  that  corresponds to a point in the
neighbourhood of the right  boundary of one of the curves in figure~\ref{fig4}.
The analysis shows that in this case, the  decay of  this  soliton  into two or
more solitons with smaller charges $\bar{Q}_{N}^{i}$ is energetically possible,
provided that $\bar{Q}_{N} = \bar{Q}_{N}^{1} + \bar{Q}_{N}^{2} + \dots$.
On the other hand, such a decay is definitely impossible if the initial $\bar{Q
}_{N}$ lies to  the  left  of  the  minimum  point  of  the $\bar{E}(\bar{Q}_{N
})/\bar{Q}_{N}$ curve.

The instability of a soliton can be either  classical  (unstable  mode  (modes)
in the spectrum  of  fluctuation)  or  quantum-mechanical (tunnelling process).
It was shown in \cite{lee_pang_1992} that a cuspidal point on a $\bar{E}(\bar{Q
}_{N})/\bar{Q}_{N}$ curve means the appearance of an unstable fluctuation mode.
Since none of the curves in figure~\ref{fig4} has a cuspidal point, we conclude
that the solitons are classically stable.
Consequently, the decay $\bar{Q}_{N} \rightarrow\bar{Q}_{N}^{1}+\bar{Q}_{N}^{2}
+ \dots$ is only possible via tunneling.
The semiclassical transition amplitude is $\propto \exp(-S_{E})$, where $S_{E}$
is an Euclidean action  of  a  field  configuration  interpolating  between the
initial and final states.
Within the framework of   the   model  \eqref{II:1},  the  semiclassical regime
is one in  which $\kappa/e^{2} \gg 1$,  $v^{2}/m \gg 1$, and $\tau = e^{2} v^{2
}/(\kappa m)$ is fixed.
In this regime, the  action $S_{E} \propto \kappa/e^{2} = v^{2}/(m \tau)\gg 1$.
Hence, the transition  amplitude  will  be  exponentially  suppressed,  and the
lifetime of the nontopological soliton can be arbitrarily large.

\subsection{The case of zero $\varepsilon$}

We now proceed to study the  case  of  $\varepsilon = 0$, which corresponds to
a first-order phase transition point.
In this point,  the  potential \eqref{II:2}  has  two  degenerate  zero minima:
symmetric at $\left\vert\phi\right \vert = 0$ and asymmetric at $\left\vert\phi
\right\vert = v$.
We shall see that the presence of the zero asymmetric minimum has a significant
effect on the properties of the soliton.
Figure~\ref{fig5} shows the $\bar{Q}_{N}(\bar{\Omega}_{\infty})$ curves for the
nontopological solitons corresponding to $\varepsilon = 0$ and different values
of $\tau$.
We see that the behaviour of the curves corresponding to nonzero $\tau$ differs
sharply from that of similar curves in figure~\ref{fig3}.
The main difference is that  for  a given $\tau$,  $\bar{Q}_{N}$, and therefore
$\bar{E}$, increases without bound  as  $\bar{\Omega}_{\infty}$  tends  to some
limiting value $\bar{\Omega}^{\lim}_{\infty}$.
The limiting   value  $\bar{\Omega}^{\lim}_{\infty}$  is  less  than  unity, it
increases with increasing  $\tau$,  and tends to unity as $\tau \rightarrow 1$.
It was found numerically that for sufficiently small $\delta  \equiv 1 - \tau$,
the limiting value $\bar{\Omega}_{\infty}^{\lim} \approx 1 - \delta/2$.
Furthermore, all the  curves  for  which  the parameter $\tau \in (0, 1)$, have
the turning point in which the derivative $d\bar{Q}_{N}/d\bar{\Omega}_{\infty}$
is infinite.
In this point, $\bar{\Omega}_{\infty}$ reaches  the  minimum value $\bar{\Omega
}^{\min}_{\infty}$, which is less than $\bar{\Omega}^{\lim}_{\infty}$.

In figure~\ref{fig5}, the behaviour of the curves  corresponding  to $\tau = 0$
or $\tau = 1$ differs from that of the other curves for which $\tau \in (0,1)$.
As in figure~\ref{fig3}, the  case  of  $\tau = 0$  corresponds  to a nongauged
two-dimensional Q-ball.
In this case, $\bar{Q}_{N}(\bar{\Omega}_{\infty})$  has  no  turning  point and
$\bar{\Omega}_{\infty}\in(\bar{\Omega}_{\infty}^{\min}, 1]$, where  the minimum
value $\bar{\Omega}_{\infty}^{\min}   =   \varepsilon \left(1 + \varepsilon^{2}
\right)^{-1/2} = 0$.
The function  $\bar{Q}_{N}(\bar{\Omega}_{\infty})$  decreases  monotonically on
the interval $(0,1)$, and it increases indefinitely $\propto\bar{\Omega}^{-3}_{
\infty}$  as $\bar{\Omega}_{\infty} \rightarrow 0$.

In the self-dual case of $\tau=1$, the soliton's energy $\bar{E}=2\pi a_{\infty
}$, and this is the minimum possible energy that a field configuration with the
magnetic flux $\bar{\Phi} =-2\pi a_{\infty}$ can have.
The soliton's  Noether  charge  $\bar{Q}_{N} = 2 \pi a_{\infty}$, and therefore
in the self-dual case,  the  derivative $d\bar{E}/d\bar{Q}_{N} = 1$.
Then, eq.~\eqref{III:29} tells us  that $\bar{\Omega}_{\infty} = 1$ for all the
self-dual nontopological solitons.
The       same      conclusion      follows      from       eq.~\eqref{III:33}.
This is consistent with  the  results presented in figure~\ref{fig5}, where the
self-dual  solitons  correspond  to  the  vertical  black  line at $\bar{\Omega
}_{\infty} = 1$.
Hence, the self-dual nontopological  solitons are parameterized in terms of the
boundary value $a_{\infty}$, whereas the non-self-dual nontopological  solitons
can be parameterized either in  terms of $\bar{\Omega}_{\infty}$ or in terms of
$a_{\infty}$, which  are  related  by  a  functional  dependency  in this case.

It was shown in ref.~\cite{jlw_prd_1990} that for the self-dual  nontopological
soliton, the  boundary value  $a_{\infty} > 2$, and its lower limit is equal to
two.
In this case, the lower limits  of  the  energy  $\bar{E}$  and  Noether charge
$\bar{Q}_{N}$ are equal to $2 \pi a_{\infty} = 4\pi$,  which is consistent with
figure~\ref{fig5}.
As $\bar{\Omega}_{\infty}  \rightarrow  1$,  the  non-self-dual  nontopological
soliton passes into the thick-wall regime.
It follows  from  figure~\ref{fig5}  that  in  this  regime, $\bar{Q}_{N}$, and
consequently $\bar{E}$, remain  finite and increase with an increase in $\tau$.
In particular, it was found  numerically  that  for  sufficiently small $\delta
\equiv 1 - \tau$, the value of $2 - a_{\infty} \propto \delta^{1/2}$, and hence
$a_{\infty} \rightarrow 2$ from  below  as $\tau$ tends to the self-dual limit.

Figure~\ref{fig6} shows the dependence  of  the  ratio $\bar{E}/\bar{Q}_{N}$ on
$\bar{Q}_{N}$   for  zero  $\varepsilon$   and  different  $\tau  \in  [0, 1]$.
We  see   that   the   behaviour   of   the   $\bar{E}/\bar{Q}_{N}$  curves  in
figure~\ref{fig6}  differs significantly from that in figure~\ref{fig4}.
In particular,   all   the    curves   in   figure~\ref{fig6},  except  for one
corresponding to  the  self-dual case of $\tau = 1$, tend from above to certain
limiting values as  $\bar{Q}_{N} \rightarrow \infty$, whereas all the curves in
figure~\ref{fig4}, except for one  corresponding to the nongauged case of $\tau
= 0$,  terminate at finite values of $\bar{Q}_{N}$.
Furthermore, unlike figure~\ref{fig4},  the  curves in figure~\ref{fig6} do not
have a minimum at the finite $\bar{Q}_{N}$.
Hence  the  decay  $\bar{Q}_{N}  \rightarrow  \bar{Q}_{N}^{1} + \bar{Q}_{N}^{2}
+ \dots$ discussed above is impossible in this case.
It is easy to see that the  limiting values of the curves in figure~\ref{fig6},
are  equal  to  the  corresponding  values   of  $\bar{\Omega}_{\infty}^{\lim}$
determined in the analysis of figure~\ref{fig5}.
Furthermore, it was found numerically that for $\tau \in (0,1)$, the difference
$\bar{E}/\bar{Q}_{N}-\bar{\Omega}_{\infty }^{\lim}$   tends  to  zero  $\propto
\bar{Q}_{N}^{-1}$ as $\bar{Q}_{N} \rightarrow \infty$.
In the nongauged case of $\tau = 0$,  the  ratio $\bar{E}/\bar{Q}_{N}$ tends to
zero $\propto \bar{Q}_{N}^{-1/3}$ as $\bar{Q}_{N} \rightarrow \infty$.
In the self-dual case of $\tau=1$,  the soliton's energy $\bar{E}=\bar{Q}_{N}$,
and therefore the ratio $\bar{E}/\bar{Q}_{N} = 1$. 
It follows that for  $\tau = 1$,  the  $\bar{E}/\bar{Q}_{N}$  curve degenerates
into the ray $\bar{E}/\bar{Q}_{N}=1$ starting from $\bar{Q}^{\min}_{N} = 4\pi$,
in accordance with figure~\ref{fig6}.

As in figure~\ref{fig4}, the  ratio  $\bar{E}/\bar{Q}_{N}$ in figure~\ref{fig6}
increases with  an  increase  in  $\tau$  at  fixed  $\bar{Q}_{N}$ resulting in
a decrease in the soliton's binding  energy $\bar{E}_{\text{b}} = \bar{Q}_{N} -
\bar{E}$.
In particular, it was found numerically that for sufficiently large $\bar{Q}_{N
}$ and $\delta \equiv 1 - \tau \ll 1$,  the  ratio $\bar{E}/\bar{Q}_{N} \approx
1 - \delta/2$.
This  is  consistent  with  the  facts  that  $\bar{E}/\bar{Q}_{N}  \rightarrow
\bar{\Omega}_{\infty}^{\lim}$   as   $\bar{Q}_{N} \rightarrow \infty$, and that
$\bar{\Omega}_{\infty}^{\lim} \approx 1 - \delta/2$ when $\delta \ll 1$.
This is also consistent  with  an  estimate  of  the energy  of quasi-self-dual
solitons \cite{bazeia_prd_1991_b}.
We recall that the parameter $\tau = m_{A}/(2 m) = m_{A}/m_{H}$ is the ratio of
the mass of a gauge boson to the mass of a Higgs boson.
Hence, in analogy with  the  standard Maxwell-Higgs system, we can say that the
parametric domain $\tau >1$ ($0< \tau < 1$) corresponds to the type-I (type-II)
region of superconductivity.
Then  it  follows  from  figures~\ref{fig1}  and  \ref{fig6}  that  the  stable
nontopological solitons exist  in  the  type-II  region  of  superconductivity,
whereas  there are no nontopological solitons in the type-I region.

\begin{figure}[tbp]
\begin{center}
\includegraphics[width=0.6\textwidth]{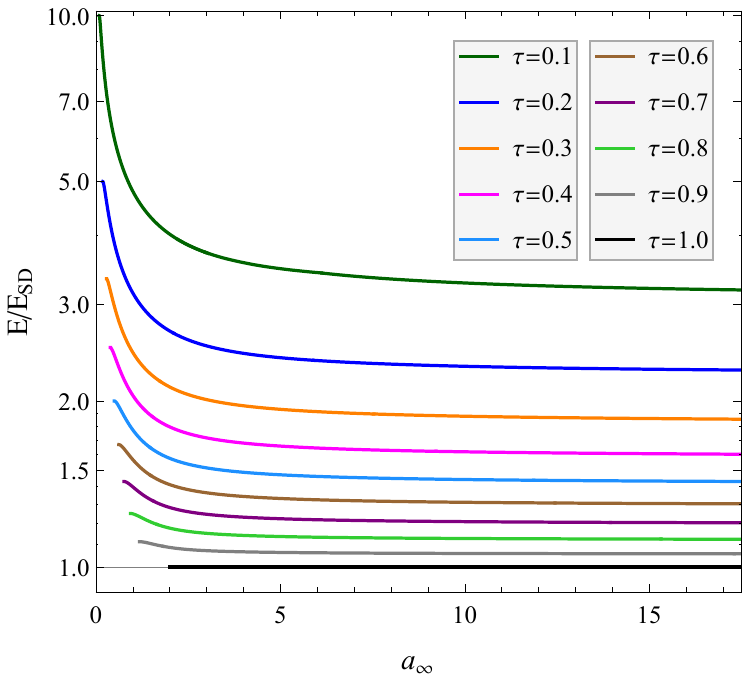}
\caption{\label{fig7}    Dependencies  of  the  ratio  of the energy $E$ of the
nontopological soliton to  the  Bogomol'nyi bound  $E_{\text{SD}} = 2 \pi v^{2}
a_{\infty}$  on  $a_{\infty}$.    The dependencies  correspond to the parameter
$\varepsilon = 0$   and   different   values  of  the  parameter  $\tau = e^{2}
v^{2}/(\kappa m)$.}
\end{center}
\end{figure}

In refs.~\cite{hong_prl_1990,  jw_prl_1990, jlw_prd_1990}, it was shown that in
the model \eqref{II:1}, the energy  of  a field configuration carrying magnetic
flux $\Phi$ cannot  be less  than $e v^{2} \left \vert \Phi \right\vert = 2 \pi
v^{2}  a_{\infty}$.
The  Bogomol'nyi bound $E_{\text{SD}}= 2 \pi v^{2} a_{\infty}$ is achieved only
in the self-dual case of $(\varepsilon, \tau) = (0, 1)$.
Let $E$ be  the  energy  of  a  nontopological  soliton  with  a given value of
$a_{\infty}$ and $E_{\text{SD}} = 2 \pi v^{2}  a_{\infty}$ be the corresponding
Bogomol'nyi bound.
Figure~\ref{fig7} shows  the dependencies of the ratio $E/E_{\text{SD}}$ on the
boundary value $a_{\infty}$.
The dependencies correspond  to  the  parameter $\varepsilon = 0$ and different
values of the parameter $\tau = e^{2}v^{2}/(\kappa m)$.
From figure~\ref{fig7}, it  follows that in the self-dual point $\tau = 1$, the
ratio $E/E_{\text{SD}} =1$ for all the values of $ a_{\infty}\in \left(2,\infty
\right)$.
For a  given  $a_{\infty}$,  the  ratio  $E/E_{\text{SD}}$  increases as $\tau$
decreases, i.e.,  as  we move away from the self-dual point $\tau = 1$.
In particular, the  ratio  $E/E_{\text{SD}}$  increases  without bound as $\tau
\rightarrow 0$.
Furthermore, in  figure~\ref{fig7}, the  ratios  $E/E_{\text{SD}}$  tend to the
certain limiting values as $a_{\infty} \rightarrow \infty$.

It can  easily  be  checked that the dimensionless Noether charge $\bar{Q}_{N}$
of the  nontopological  soliton  can  be  written  as  $2 \pi a_{\infty}/\tau$.
It follows  that  the  ratio  $E/E_{\text{SD}} = \bar{E}/\left(2 \pi a_{\infty}
\right) = \tau^{-1}\bar{E}/\bar{Q}_{N}$.
In turn, it has been  established  above  that  the ratio $\bar{E}/\bar{Q}_{N}$
tends to the  limiting  value $\bar{\Omega}^{\lim}_{\infty}$   as  $\bar{Q}_{N}
\rightarrow \infty$.
Combining the last two results,  we  conclude that for a given value of $\tau$,
the limiting value of the corresponding curve in figure~\ref{fig7}  is equal to
$\bar{\Omega}^{\lim}_{\infty}/\tau$.
In the vicinity of the self-dual point, the difference $\delta = 1-\tau \ll 1$,
and the limiting  value $\bar{\Omega}^{\lim}_{\infty}/\tau\approx 1+ \delta/2$,
where we use the fact that $\bar{\Omega}_{\infty}^{\lim}  \approx 1 - \delta/2$
when $\delta \ll 1$.

From figure~\ref{fig6}, it follows  that  for sufficiently large $\bar{Q}_{N}$,
the ratio $\bar{E}/\bar{Q}_{N}$ is practically independent of $\bar{Q}_{N}$ and
decreases with a decrease in $\tau$.
In turn,  from  figure~\ref{fig7},  it  follows  that  for  sufficiently  large
$a_{\infty}$, the ratio $\bar{E}/\bar{E}_{\text{SD}}$ is practically independent
of $a_{\infty}$.
However,  unlike  figure~\ref{fig6},  the  ratio  $\bar{E}/\bar{E}_{\text{SD}}$
increases with a decrease in $\tau$.
This difference in the behaviour of the ratios $\bar{E}/\bar{Q}_{N}$ and $\bar{E
}/\bar{E}_{\text{SD}}$ is due to the fact that they are related by the equation
$\bar{E}/\bar{Q}_{N} = \tau \bar{E}/\bar{E}_{\text{SD}}$.
As $\tau$  decreases,  the ratio  $\bar{E}/\bar{E}_{\text{SD}}$  increases, but
this increase is outweighed by the decrease in the multiplier $\tau$.
As a result, the  ratio  $\bar{E}/\bar{Q}_{N}$  decreases  with  a  decrease in
$\tau$, in accordance with figure~\ref{fig6}.

We conclude this section by  explaining  the difference in the behaviour of the
curves in figures \ref{fig3} and \ref{fig5}. 
In particular, we shall show that the  energy  and charge of the nontopological
soliton can  be  arbitrarily  large  only  if  the parameter $\varepsilon = 0$.
It is evident that with the unlimited  increase  in  the energy (charge) of the
soliton, its radial size also increases unlimitedly.
For this to be possible, the  amplitude of  the  scalar field  $\left\vert \phi
\right\vert = v f(r)$, and  hence  the  ansatz  function $f(r)$, must be nearly
constant over an indefinitely large interval of $r$.
Eq.~\eqref{III:3a} tells us  that  for  this  to be possible, the initial value
$f_{0}$ must satisfy the  condition $\bigl. \partial \tilde{V}/\partial f \bigr
\vert_{f=f_{0}} \approx 0$, i.e., $f_{0}$  must be in the close vicinity of the
local minimum of the potential \eqref{III:4}.
Furthermore,  both  $\Omega(r)$  and  $a(r)/r$  must  be arbitrarily small over
an indefinitely large interval of $r$.
According to eqs.~\eqref{III:14b}  and \eqref{III:14c},  the last condition can
only be satisfied if the initial value $\Omega_{0}$ is arbitrarily small.

The local minimum of the potential is located  at $f_{\min} = 3^{-1/2}\bigl(2 +
\left(1-3\varepsilon^{2}\right)^{1/2}\bigr)^{1/2}$, provided that the parameter
$\varepsilon \in \left[0, 3^{-1/2}\right)$.
Hence,  the  soliton  of  arbitrarily  large  energy and charge cannot exist if
$\varepsilon \ge 3^{-1/2}$, since  there  is  no local minimum of the potential
\eqref{III:4} in this case.
Further, the  virial  relation \eqref{III:31} tells us that the kinetic part of
the soliton's energy $E_{T} = 2 \pi v^{2} \int_{0}^{\infty}\Omega ^{2}f^{2}rdr$
must be equal to the potential part $E_{P}=2\pi v^{2}\int_{0}^{\infty}\tilde{V}
\left( f\right) rdr$.
Let us suppose that the  solitons  of  arbitrarily  large energy and charge may
exist in the general case of nonzero $\varepsilon$ less than $3^{-1/2}$.
In this case, the value of  $\tilde{V}\left(f_{\min}\right)$ and, consequently,
of $\tilde{V}\left(f\right)$  is  not  arbitrarily  small, whereas the value of
$\Omega^{2} f^{2}$  is   arbitrarily    small,   since   the   ansatz  function
$\Omega(r)$ must  be  arbitrarily  small over an indefinitely large interval of
$r$.
It follows that the virial relation $E_{T} =E_{P}$ cannot be met, and therefore
the solitons  of  arbitrarily  large  energy  and  charge do not exist when the
parameter $\varepsilon \in \left(0, 3^{-1/2}\right)$.
The   situation   changes   when    the   parameter   $\varepsilon$   vanishes.
In this  case, the  potential  \eqref{III:4}  has  two  degenerate  zero minima
located at $f = 0$ and $f = 1$, respectively.
With the increase in the  energy  and  charge of the soliton, the initial value
$f_{0} \rightarrow 1$, and therefore the value of $\tilde{V}(f_{0}) \rightarrow
0$.
In this case, $\tilde{V}\left(f\right)$  and $\Omega^{2} f^{2}$ are of the same
order of smallness, and  therefore  the  virial relation $E_{T} = E_{P}$ can be
satisfied.
This means that the  nontopological  solitons  of  arbitrarily large energy and
charge can exist when the parameter $\varepsilon = 0$.

\section{Conclusion}                                             \label{sec:VI}

In the present paper,  we  have  studied   a   nontopological   soliton  of the
Chern-Simons-Higgs gauge model.
The general non-self-dual case  was  considered  in  which the soliton's energy
exceeds the Bogomol'nyi bound.
The properties of the soliton  were  studied  using  analytical  and  numerical
methods.
In particular, it was shown that  the  nontopological soliton is an extremum of
the energy functional at a fixed Noether (electric) charge.
This basic property  of  the  soliton  results  in  the  important differential
relation    \eqref{III:29}    and     its    consequences   \eqref{III:30a}  --
\eqref{III:30c}.
Another important property is the virial relation~\eqref{III:31}, from which it
follows that  the  kinetic  part   of  the  soliton's  energy  is  equal to its
potential part.

Dimensional analysis  reveals  that  the  nontopological  soliton  solution  is
essentially determined  by  the  values  of the three  dimensionless parameters
$\varepsilon$, $\tau$, and $\bar{\Omega}_{\infty}$.
Using  numerical  methods,  we  determined  the  domain  of  existence  of  the
non-self-dual soliton solution in the $\varepsilon \tau$-plane.
In particular, we established that at fixed $\varepsilon$, the soliton solution
exists only if $0 \le \tau < \tau_{\max}(\varepsilon) \le 1$.
The case of $\tau = 0$  corresponds  to  the  two-dimensional  nongauged Q-ball
\cite{lee_pang_1992}.
The self-dual  nontopological   soliton   considered   in   \cite{jlw_prd_1990}
corresponds to the point $(\varepsilon, \tau) = (0, 1)$  lying  on the boundary
of the domain of existence.
The non-self-dual solitons are absolutely stable against emission of elementary
particles,  whereas  the  self-dual  solitons  are  just  at  the  threshold of
stability. 

For a fixed $\tau \in \left(0, 1\right)$,  the properties of the nontopological
soliton depend significantly on whether the parameter $\varepsilon$ is equal to
zero or not.
If the parameter $\varepsilon \ne 0$, then the energy $\bar{E}$ and the Noether
charge $\bar{Q}_{N}$ of the soliton cannot be arbitrarily large.
In this  case,  the  curves  $\bar{E}(\bar{\Omega}_{\infty})$  and  $\bar{Q}_{N
}(\bar{\Omega}_{\infty})$  have  turning  points  at some $\bar{\Omega}_{\infty
}^{\min} < 1$ and reach  finite  maximum values at $\bar{\Omega}_{\infty} = 1$.
Such a behaviour of the curves $\bar{E}(\bar{\Omega}_{\infty})$ and $\bar{Q}_{N
}(\bar{\Omega}_{\infty})$ is similar to  that  of  the corresponding curves for
a  three-dimensional   electrically   charged   Q-ball  \cite{gulamov_prd_2015,
 loginov_prd_2020}.

In contrast, the  energy  and  the  Noether  charge  of  the  soliton  can take
arbitrarily large values, provided that the parameter $\varepsilon = 0$.
In this case, the  curves   $\bar{E}(\bar{\Omega}_{\infty})$   and  $\bar{Q}_{N
}(\bar{\Omega}_{\infty})$  also  have  turning   points  at  some  $\bar{\Omega
}_{\infty}^{\min} < 1$.
However, unlike the previous case, now both curves have a vertical asymptote at
some $\bar{\Omega}^{\lim}_{\infty}$  such   that  $\bar{\Omega}^{\min}_{\infty}
< \bar{\Omega}^{\lim}_{\infty} < 1$.
As $\bar{\Omega}_{\infty}$  tends  to  $\bar{\Omega}^{\lim}_{\infty}$  from the
left, the energy and charge of  the soliton increase without bound, while their
ratio tends to $\bar{\Omega}^{\lim}_{\infty}$.
The magnitude of the magnetic flux of the soliton and its angular momentum also
increase without bound as $\bar{\Omega}_{\infty} \rightarrow \bar{\Omega}^{\lim
}_{\infty}$.
The value of $\bar{\Omega}^{\lim}_{\infty}$  depends  on  $\tau$;  it increases
monotonically  from  zero  to  one  as $\tau$ increases within the same limits.

In the present work,  we  considered the nontopological  soliton  solution that
does  not depend on the azimuthal angle.
In addition  to  this solution, the  model~\eqref{II:1} also has nontopological
soliton solutions  of  the  vortex type  \cite{jlw_prd_1990}  with  the angular
dependence $\exp(i n \theta)$.
Similar to the soliton considered here, these nontopological vortices can exist
both at  zero  and  nonzero  values  of  $\varepsilon$,  i.e.,  above  and at a
first-order phase transition point.
It would  be  interesting  to  establish  the  regions  of  existence  for  the
nontopological vortices  with  different  winding  numbers $n$  by analogy with
figure~\ref{fig1}.
In addition to these   two   types  of  nontopological   soliton solutions, the
model~\eqref{II:1} also has  topological  vortex solutions \cite{hong_prl_1990,
 jw_prl_1990, jlw_prd_1990}.
However, unlike the nontopological solitons, the topological vortices can exist
below and at the first-order phase transition point.

The scalar field $\phi$  of  the  considered  nontopological  soliton  does not
vanish at finite $r$, since the ansatz function $f(r)$ has no nodes.
At the same  time,  there  exist   radially   excited   nontopological solitons
\cite{fried_prd_1976,  volkov_prd_2002,  kleihaus_prd_2005, mai_prd_2012} whose
ansatz functions have one or more nodes.
It is  obvious  that  within  the  framework  of  the  model~\eqref{II:1}, such
radially excited  solutions  cannot  be  self-dual, since their energies exceed
the Bogomol'nyi bound.
However, the existence of radially excited solutions is possible in the general
non-self-dual case.
This       question          will        be        considered        elsewhere.

\appendix
\section{The thick-wall regime of the soliton}

In figures \ref{fig3} and  \ref{fig5}, the curves $Q_{N}(\Omega_{\infty})$ that
correspond to the  non-self-dual  nontopological  solitons consist of the upper
and lower branches connected at the turning points.
The thick-wall regime  corresponds  to  those  parts  of the lower branches for
which the parameter  $\Omega_{\infty} \in \left(m - \delta m, m \right)$, where
$\delta m/m \ll 1$.
Numerical studies reveal that  in  the thick-wall regime, when $\Omega_{\infty}
\rightarrow m$, the soliton scalar field $\phi$  spreads  out in space, and its
amplitude $f$ tends  to  zero  $\propto \Delta$,  where $\Delta = \left(m^{2} -
\Omega_{\infty}^{2}\right)^{1/2}$.
From eqs.~\eqref{III:15} -- \eqref{III:17},  it  follows that the scaled radial
variable $\varrho = \Delta r$ is natural in thick-wall regime.
The transition  to   the   new  variable  $\varrho$  in  eqs.~\eqref{III:3a} --
\eqref{III:3c}, taking into account that the ansatz function $f\propto \Delta$,
and the self-consistency  of  eqs.~\eqref{III:3a} -- \eqref{III:3c}  lead us to
the conclusion that  in  the  thick-wall  regime,  the ansatz functions $a_{0}$
and $a$ are proportional to $\Delta^{2}$ and $\Delta^{0}$, respectively.
Numerical  results   confirm    such    behaviour    of    $a_{0}$    and  $a$.

In view  of  the  above,  we  transition  to the new radial variable $\varrho =
\Delta r$ and rescaled  ansatz  functions according to the formulas
\begin{equation}
f\left( r\right) =\Delta m^{-1}\tilde{f}\left( \varrho \right),\quad
a_{0}\left( r\right) =\Delta ^{2}m^{-2}\tilde{a}_{0}\left( \varrho
\right),\quad a(r)=\tilde{a}\left( \varrho \right).                 \label{A:1}
\end{equation}
Using eq.~\eqref{A:1}, we find  that  up  to  terms  of order $\Delta^{4}$, the
Lagrangian \eqref{III:25} can be written as
\begin{equation}
L=\Delta^{2}L_{2} + \Delta^{4}L_{4},                                \label{A:2}
\end{equation}
where
\begin{equation}
L_{2}=2\pi \frac{v^{2}}{m^{2}}\int\nolimits_{0}^{\infty }\left[ \frac{\kappa }{
2ev^{2}}\frac{1}{\varrho }\left( \tilde{a}_{0}\tilde{a}^{\prime }-
\tilde{a}_{0}^{\prime }\tilde{a}\right) -\tilde{f}^{\prime 2}-2
\frac{e}{m}\tilde{a}_{0}\tilde{f}^{2}-\frac{\tilde{a}^{2}}{
\varrho ^{2}}\tilde{f}^{2}-\tilde{f}^{2}+\frac{2\tilde{f}^{4}}{
1+\varepsilon ^{2}}\right] \varrho d\varrho                         \label{A:3}
\end{equation}
and
\begin{equation}
L_{4}=2\pi \frac{v^{2}}{m^{4}}\int\nolimits_{0}^{\infty }\left[\frac{e}{m}
\tilde{a}_{0}\left(1+\frac{e}{m}\tilde{a}_{0}\right) \tilde{f}^{2}-
\frac{\tilde{f}^{6}}{1+\varepsilon^{2}}\right]\varrho d\varrho.     \label{A:4}
\end{equation}
do not depend on $\Omega_{\infty}$. 

From eqs.~(\ref{III:27})  and   (\ref{III:29}),  it  follows  that  the  energy
$E(Q_{N})$  and the Lagrangian $L(\Omega_{\infty})$  of the soliton are related
through the Legendre transformation $E\left(Q_{N}\right)= \Omega_{\infty} Q_{N}
-L\left(\Omega_{\infty} \right)$, where $\Omega_{\infty} = dE/dQ_{N}$.
Using  the  known  properties   of   the   Legendre  transformation,  we obtain
successively the  expressions  of  the  Noether  charge  and  the energy of the
soliton   in     the     thick-wall     approximation   \cite{kusenko_plb_1997,
 mutamaki_npb_2000, paccetti_epjc_2001}
\begin{equation}
Q_{N}\left( \Omega _{\infty }\right) =\frac{dL}{d\Omega _{\infty }}
=-2L_{2}\Omega _{\infty }-4L_{4}\Omega _{\infty }\Delta ^{2}        \label{A:5}
\end{equation}
and $E\left(Q_{N}\right) = E\left( \Omega_{\infty}\left(Q_{N} \right) \right)$,
where
\begin{equation}
E\left( \Omega _{\infty }\right)=\Omega_{\infty }\frac{dL}{d\Omega
_{\infty }}-L\left(\Omega_{\infty}\right)=-L_{2}\left( 2\Omega_{\infty
}^{2}+\Delta^{2}\right) -L_{4}\left( 4\Omega_{\infty}^{2}\Delta
^{2}+\Delta^{4}\right).                                             \label{A:6}
\end{equation}

Let us  introduce  the  notations $Q_{N}^{\text{tw}} \equiv Q_{N}(m)$ and $\eta
\equiv  \left. dQ_{N}/d\Omega_{\infty }\right\vert _{\Omega _{\infty }  =  m}$.
From eq.~\eqref{A:5}, it follows that
\begin{equation}
Q_{N}^{\text{tw}}=-2L_{2}m \quad \text{and} \quad \eta = 8 m^{2}L_{4}-2L_{2}.
                                                                    \label{A:7}
\end{equation}
Eqs.~\eqref{A:5} -- \eqref{A:7} lead  us  to  the expressions for the soliton's
energy $E$ and the ratio $E/(m Q_{N})$ in  terms  of the Noether charge $Q_{N}$
\begin{equation}
E\approx mQ_{N}^{\text{tw}} + m\left(Q_{N} - Q_{N}^{\text{tw}}\right) +
\frac{1}{2\eta}\left(Q_{N}  -Q_{N}^{\text{tw}}\right)^{2},          \label{A:8}
\end{equation}
\begin{equation}
E/\left(m Q_{N}\right) \approx 1 + \frac{1}{2 m \eta}
\left(Q_{N}/Q_{N}^{\text{tw}}-1\right)^{2}.                         \label{A:9}
\end{equation}

It follows from figures \ref{fig3}  and  \ref{fig5}  that the lower branches of
the curves  $Q_{N}(\Omega_{\infty})$   that  correspond  to  the  non-self-dual
nontopological solitons  are  decreasing  functions  of  $\Omega_{\infty}$, and
hence $\eta$ is negative for them.
Then, eq.~\eqref{A:9}  tells  us  that  the  ratio  $E/(m Q_{N})$  must reach a
maximum equal to $1$ at $Q_{N} =Q_{N}^{\text{tw}}$, i.e., at the limiting point
of the thick-wall regime.
Figures \ref{fig4}  and  \ref{fig6}  show   that   this   is  indeed  the case.

\acknowledgments

This  work  was  supported   by   the   Russian  Science  Foundation,  grant No
23-11-00002-Ext.




\end{document}